\documentclass[fleqn,usenatbib,onecolumn]{mnras}
\usepackage{newtxtext,newtxmath}
\usepackage[T1]{fontenc}
\usepackage{ae,aecompl}

\usepackage{multicol}
\usepackage{graphicx}	
\usepackage{amsmath}
\usepackage{hyperref}
\usepackage[dvipsnames]{xcolor}

\definecolor{mnrasblue}{rgb}{0.07, 0.39, 0.8}
\hypersetup{
    colorlinks=true,
    linkcolor=mnrasblue,      
    urlcolor=mnrasblue,
    citecolor=mnrasblue
}

\newcommand{\bs}{\boldsymbol}
\newcommand{\aperp}[1]{{\bs #1}_\perp}
\newcommand{\apar}[1]{{#1}_\parallel}
\newcommand{\kunit}{h\,{\rm Mpc}^{-1}}
\newcommand{\kFG}{\apar{k}^{\rm FG}}

\title[Clustering redshifts with 21-cm bispectra]{Clustering redshifts with the 21cm-galaxy cross-bispectrum}

\author[C. Guandalin {\sl et al}.]{
Caroline Guandalin,$^{1}$\thanks{E-mail: caroline.guandalin@usp.br}
Isabella P. Carucci,$^{2,3}$
David Alonso,$^{4}$
Kavilan Moodley$^{5,6}$
\\
$^{1}$Departamento de F\'isica Matem\'atica, Instituto de F\'isica, Universidade de S\~ao Paulo, Rua do Mat\~ao 1371, CEP 05508-090, S\~ao Paulo, Brazil\\
$^{2}$Dipartimento di Fisica, Universit\`a degli Studi di Torino, via P.\ Giuria 1, 10125 Torino, Italy\\
$^{3}$INFN -- Istituto Nazionale di Fisica Nucleare, Sezione di Torino, via P.\ Giuria 1, 10125 Torino, Italy\\
$^{4}$ Department of Physics, University of Oxford, Denys Wilkinson Building, Keble Road, Oxford OX1 3RH, UK \\
$^{5}$ Astrophysics Research Centre, University of KwaZulu-Natal, Westville Campus, Durban 4041, South Africa \\
$^{6}$ School of Mathematics, Statistics \& Computer Science, University of KwaZulu-Natal, Westville Campus, Durban4041, South Africa
}

\date{}

\pubyear{2021}

\begin{document}
\label{firstpage}
\pagerange{\pageref{firstpage}--\pageref{lastpage}}
\maketitle

\begin{abstract}
The cross-correlation between 21-cm intensity mapping experiments and photometric surveys of galaxies (or any other cosmological tracer with a broad radial kernel) is severely degraded by the loss of long-wavelength radial modes due to Galactic foreground contamination. Higher-order correlators are able to restore some of these modes due to the non-linear coupling between them and the local small-scale clustering induced by gravitational collapse. We explore the possibility of recovering information from the bispectrum between a photometric galaxy sample and an intensity mapping experiment, in the context of the clustering-redshifts technique. We demonstrate that the bispectrum is able to calibrate the redshift distribution of the photometric sample to the required accuracy of future experiments such as the Rubin Observatory, using future single-dish and interferometric 21-cm observations, in situations where the two-point function is not able to do so due to foreground contamination. We also show how this calibration is affected by the photometric redshift width $\sigma_{z,0}$ and maximum scale $k_{\rm max}$. We find that it is important to reach scales $k \gtrsim 0.3\,\kunit$, with the constraints saturating at around $k\sim 1\,\kunit$ for next-generation experiments.
\end{abstract}


\section{Introduction}
Galaxy redshift surveys are a cornerstone of contemporary cosmology, with the spectroscopic ones providing highly accurate maps of the three-dimensional galaxy distribution \citep{1312.5490,1611.00036,2007.08991}. When dealing with discrete dark matter tracers, like galaxies, it is imperative to have a considerable number of observations for the $n$-point signal to overcome its noise. With the spectroscopic technique, acquiring millions of galaxy spectra is a time consuming process and, even though it is desirable to have high-precision redshifts to reach specific science goals -- e.g. baryon acoustic oscillations \citep{astro-ph/0701079}, redshift-space distortions \citep{1987MNRAS.227....1K}, or Doppler lensing \citep{bacon2014} --, interesting physical signatures coming from relativistic and primordial non-Gaussian effects require us to observe the largest scales that are only reachable with a large volume coverage \citep{0710.4560,0807.1770,1305.6928,1505.07596,1507.03550,1710.09465}.

Photometric surveys appear as a promising, less expensive route to map the 3D distribution of matter inhomogeneities, covering vast regions of the sky while sampling billions of galaxies in a short amount of time. However, this comes at the expense of large redshift uncertainties, resulting in a limited radial resolution. Specifically, the presence of photometric-redshift (photo-$z$) errors damp small radial scales, acting similarly to redshift-space distortions \citep{blake2005}. While it may be possible to obtain relatively precise photo-$z$s for galaxy samples featuring strong spectral features (e.g. the 4000 \AA \, break), this is not the case for the overall galaxy population, resulting in a severe loss of high-$\apar{k}$ modes\footnote{In what follows we will label radial (angular) scales with a $\parallel$ ($\perp$) subscript.}. The presence of large redshift uncertainties impact summary statistics used in cosmological inference and, eventually, degrade cosmological parameter constraints \citep{ross2011,montero2018}.

Fortunately, the presence of different experiments tracing the same large-scale structure (LSS) opens up a range of possibilities to explore combinations of multiple tracers that can complement each other to overcome the specific barriers of each survey. In particular, an avenue to diminish uncertainties in the redshift distribution of a photometric sample, and their impact on cosmological analyses, is to cross-correlate the photometric sample with a spectroscopic one, in what is commonly called the ``clustering redshifts'' technique \citep{newman2008, mcquinn2013}. The limiting factor in this case is the overlap in area, redshift, and range of scales probed by both surveys.

For many LSS analyses, resolving individual galaxies to obtain a detailed map at small scales, as is done spectroscopic galaxy surveys (in the optical or in HI), is not necessary. In the intensity mapping (IM) approach, a map of the LSS is built from the integrated 21-cm emission line of many (unresolved) individual sources at a given angular scale, providing a fast way to map large volumes, up to high redshifts, with spectroscopic precision \citep{astro-ph/0401340,astro-ph/0512263,0709.3672,0708.3392,0902.3091,0908.3796,1209.0343,1208.0331,1304.3712,1405.1452,2102.04946}. Therefore, it might be possible, in principle, to make use of IM data to calibrate photometric redshifts \citep{alonso2017, 2019MNRAS.482.3341C}.

Although the IM technique overcomes the weakness of the neutral hydrogen signal of an individual galaxy, by integrating the emission from many galaxies and, thus, delivering a larger signal-to-noise, it is prone to foreground contamination that surpasses the strength of the cosmological signal by several orders of magnitude \citep{santos2005,wolz2013,alonso2014,2019MNRAS.488.5452C,2020MNRAS.499..304C,spinelli2021}. Foreground contamination is dominant on the largest radial scales, and its removal renders only the smaller (high $\apar{k}$) modes of the cosmological signal accessible for analysis. This has a severe impact on the cross-correlation between IM and photometric survey data and, therefore, on the possibility of applying the standard clustering redshifts method.

In this paper we will explore a method to overcome this difficulty, by employing higher-order correlators encoding the non-linear coupling between long- and short-wavelength modes induced by gravity. The idea is based on the principle that we can access the lost radial modes in the IM experiment by exploring the impact of the long-wavelength fluctuations on the local HI inhomogeneities. In particular, we will focus on the ability of the bispectrum involving one photometric galaxy sample, which contains the long-wavelength information, and two IM modes to reconstruct the redshift distributions of the photometric sample. In spirit, this is similar to the idea of tidal-mode reconstruction applied to the cross-correlation of 21-cm data with CMB lensing and the projected kinematic Sunyaev-Zel'dovich effect \citep{li2018,hotinli2021}. 

This paper is structured as follows. The rationale behind this method is presented in Section \ref{sec:clustz}, where we derive expressions for the uncertainty on photo-$z$ parameters achievable through clustering redshifts with the power spectrum and the bispectrum. The survey details and specifications, along with the LSS modelling used, are described in Section \ref{sec:surveys}. Our main results are presented and discussed in Section \ref{sec:results}, where we quantify the regime where the bispectrum constraints overcome the power spectrum, as well as the photo-$z$ constraints achievable by different Stage-IV surveys. Finally, conclusions are drawn in Section \ref{sec:conclusions}. The relevant mathematical derivations are presented in Appendix \ref{app:math}, and some variations of the results presented in Section \ref{ssec:simplified_results} are shown and discussed in Appendix \ref{ap:extra}.

\vspace{-0.3cm}
\begingroup
\allowdisplaybreaks
\section{Clustering redshifts with 2- and 3-point functions}\label{sec:clustz}
  In the absence of precise redshift information, LSS analyses are carried out tomographically, projecting the data from different galaxy samples (usually lying on a more or less compact range of redshifts) onto the two-dimensional celestial sphere. In the flat-sky approximation (which we use throughout this paper), let $\Delta_g(\aperp{r})$ be the projected galaxy overdensity in a given redshift bin, related to the three-dimensional overdensity $\delta_g(\apar{r},\aperp{r})$ via
  \begin{equation}
    \Delta_g(\aperp{r})=\int {\rm d}\apar{r}\,\phi(\apar{r})\,\delta_g(\apar{r},\aperp{r}),\label{eq:projection}
  \end{equation}
  where $\phi(\apar{r})$ is the radial selection function, which is unknown a priori. The object of the ``clustering redshifts'' or ``cross-correlation redshifts'' method is to constrain $\phi(\apar{r})$ through the use of a second tracer of the same underlying LSS, but with a well-understood radial distribution. This second tracer is commonly a spectroscopic galaxy sample, although other probes, such as CMB lensing \citep{2021MNRAS.502..876A,2021A&A...649A.146R}, may be used to recover certain properties of $\phi(\apar{r})$.

  The standard implementation of the clustering redshifts approach involves the use of the two-point cross-correlation between the projected overdensity and ``spectroscopic'' tracer. Treating $\phi(\apar{r})$ as a histogram (see Eq. \ref{eq:decomposition}), the cross-correlation of $\Delta_g(\aperp{r})$ with spectroscopic sources within one of the histogram bins $a$ will be proportional to the number of photometric galaxies in that bin (and zero if there are none). Thus, if the linear biases of the spectroscopic and photometric samples can be measured or calibrated, all histogram bin heights $\phi_a$ characterising $\phi(\apar{r})$ can be determined from these cross-correlations.

  This simple picture becomes murkier when we use 21-cm intensity mapping observations as the ``spectroscopic'' tracer. In this case, although the 21-cm data are able to resolve the smaller radial scales, long-wavelength radial modes are lost due to foreground contamination. Thus, since the projected galaxy overdensity $\Delta_g$ is only sensitive to radial scales larger than the support of $\phi(\apar{r})$, the range of scales over which both tracers have a significant cross-correlation is severely reduced, degrading significantly the performance of the standard clustering redshifts approach \citep{alonso2017,2019MNRAS.485.5519W,2019MNRAS.482.3341C,2019MNRAS.488.5452C}. The use of higher-order correlators, exploiting the coupling between small-scale and large-scale modes caused by non-linear gravitational collapse, is a potential workaround to this problem.

  The next two subsections present a Fisher matrix-based approach to calculate the uncertainties on $\phi(\apar{r})$ achievable through the use of 2-point and 3-point correlators involving a photometric galaxy sample and a set of HI maps. The details of this calculation are described in Appendix \ref{app:math}, and we only provide the final results here. For simplicity, and since it will allow us to connect the final constraints with the specific requirements for photo-$z$ calibration in Stage-IV experiments, we will assume that the radial selection function is described by a Gaussian function centred at distance $\apar{r}^*$ with standard deviation $\apar{\sigma}$:
  \begin{equation}\label{eq:phi_gaussian}
    \phi(\apar{r})\propto\exp\left[-\frac{(\apar{r}-\apar{r}^*)^2}{2\apar{\sigma}^2}\right].
  \end{equation}
  \endgroup

  \subsection{Clustering redshifts with two-point functions}\label{ssec:clustz.2pt}
    The Fisher matrix for a data vector made up of a set of power spectra between two-dimensional fields ${\cal P}^{XY}(k_\perp)$ is given by:
    \begin{equation}\label{eq:FisherP}
      F^P_{\alpha\beta}=\sum_{{\bs X}{\bs X}'}\frac{A}{4\pi}\int_0^\infty {\rm d}k_\perp\,k_\perp\,\partial_\alpha {\cal P}^{XY}(k_\perp)\,\partial_\beta {\cal P}^{X'Y'}(k_\perp)\,{\cal I}^{XX'}(k_\perp)\,{\cal I}^{YY'}(k_\perp),
    \end{equation}
    where $A$ is the comoving survey area, ${\cal I}^{XY}$ is the $XY$ element of the inverse matrix of ${\cal P}^{XY}$, ${\bs X} = \lbrace X, Y\rbrace$ runs over all the maps used, and $\partial_\alpha$ denotes the derivative with respect to a given parameter $\theta_\alpha$. In our case, this will correspond to any parameters describing the radial selection function, such as a set of histogram heights, Fourier coefficients or the Gaussian parameters $(\apar{r}^*, \apar{\sigma})$. Once $F^P_{\alpha\beta}$ has been calculated, the covariance of all parameters is simply given by its inverse.

    Following the derivation in Appendix \ref{app:math}, the variances of $\apar{r}^*$ and $\apar{\sigma}$ can be approximated as:
    \begin{align}\label{eq:var_r_P}
      {\rm Var}^{-1}_P(\apar{r}^*)&\simeq \frac{A}{2\pi}\int {\rm d}k_\perp k_\perp \, {\cal I}^{TT}(k_\perp)\int \frac{{\rm d}\apar{k}}{2\pi}\apar{k}^2{\rm e}^{-\apar{k}^2\apar{\sigma}^2}\frac{P_{gh}^2(\apar{k},k_\perp)}{P_{hh}(\apar{k},k_\perp)},\\\label{eq:var_s_P}
      {\rm Var}^{-1}_P(\apar{\sigma})&\simeq \frac{A}{2\pi}\int {\rm d}k_\perp k_\perp \, {\cal I}^{TT}(k_\perp)\int \frac{{\rm d}\apar{k}}{2\pi}\apar{\sigma}^2\apar{k}^4{\rm e}^{-\apar{k}^2\apar{\sigma}^2}\frac{P_{gh}^2(\apar{k},k_\perp)}{P_{hh}(\apar{k},k_\perp)},
   \end{align}
   and their covariance is ${\rm Cov}(\apar{r}^*,\apar{\sigma})=0$. Here $P_{gg}$, $P_{gh}$ and $P_{hh}$ are the auto- and cross-spectra of the galaxy ($g$) and HI ($h$) fields in three dimensions. ${\cal I}^{TT}$ is the element of ${\cal I}^{XY}$ corresponding to the autocorrelation of the projected photometric sample, and is given by
   \begin{equation}\label{eq:ITT_main}
     {\cal I}^{TT}(\aperp{k})=\left\{\int \frac{{\rm d}\apar{k}}{2\pi}\,{\rm e}^{-\apar{k}^2\apar{\sigma}^2} P_{gg}(k)\left[1-\frac{P^2_{gh}(k)}{P_{gg}(k)P_{hh}(k)}\right]\right\}^{-1},
   \end{equation}
   where we use $k$ as a shorthand for $(\apar{k},\aperp{k})$. Note that, implicitly, all autocorrelations ($P_{gg}$ and $P_{hh}$) include noise contributions, as described in Section \ref{sec:surveys}.
   
   A number of approximations have been made to reach this result:
   \begin{enumerate}
     \item We have only considered the Gaussian contribution to the power-spectrum covariance (4-point function). This likely not a bad approximation on the scales used here, and in the presence of instrumental noise.
     \item We have disregarded all information from the autocorrelation of the projected galaxy overdensity field $\Delta_g(\aperp{k})$. As discussed in Appendix \ref{ssec:fisher}, this is appropriate for $\apar{r}^*$, but the discarded information is, at face value, important for $\apar{\sigma}$. This is because the amplitude of the projected galaxy autocorrelation is highly sensitive to the width of the redshift distribution \citep{2021MNRAS.502..876A}. In a realistic scenario this sensitivity would be degenerate with the unknown galaxy bias (which we do not marginalise over here). Furthermore, for the purposes of this exercise, we are only interested in the contribution from the 21-cm cross-correlation.
     \item We have assumed the galaxy shot noise to be described by a Poisson distribution, and that its stochasticity is uncorrelated with possible small-scale fluctuations for the HI samples. Furthermore, because the intensity mapping observations are not built from discrete tracers, but from the integrated emission of neutral hydrogen inside galaxies, a Poissonian noise should be negligible if compared with the HI autocorrelation signal. Nonetheless, if the number of HI-emitting galaxies is low, such that the brightness temperature is not enough to deliver a high signal-to-noise, then corrections should be accounted for \citep{wolz2016,wolz2019}.
     \item We have only kept the leading-order terms in $\epsilon=P_{gh}/P_{hh}$. This is exact in the limit of zero mode overlap between the HI maps and the projected galaxy distribution, and would lead to at most $\mathcal{O}(1)$ corrections\footnote{By $\mathcal{O}(1)$ we mean corrections that could change our results by a factor of $\sim2$. Similarly, the inclusion of off-diagonal contributions in the covariance matrix, at least for the coupling of squeezed bispectrum configurations, have also been shown to degrade primordial non-Gaussian constraints by a factor of 2 \citep{biagetti2021}.} in the final result (similar to the effect of other approximations used here). As mentioned in Appendix \ref{app:gaussian} and item (ii) above, the effects of this assumption on the final constraints on $\apar{r}^*$ are negligible.
   \end{enumerate}

   Eq. \ref{eq:var_r_P} is illustrative of the problem posed by foregrounds to the clustering redshifts method. Focusing on the inner integral over $\apar{k}$, we see that, on the one hand, the exponential prefactor dampens the contribution of radial modes smaller than the photo-$z$ width $\apar{\sigma}$. On the other hand, the ratio $P_{gh}/P_{hh}$ is only significant on scales smaller than those dominated by foreground contamination. If  $\sigma_\parallel$ is larger than the largest foreground-clean scale, both factors conspire to suppress the integrand over the whole $\apar{k}$ range. Radio foregrounds are expected to affect 21-cm observations at least on scales $\apar{k}\lesssim \kFG\sim 0.02\,\kunit$ \citep{2015PhRvD..91h3514S}, which coincides with the physical scale corresponding to a photo-$z$ width $\sigma_z\simeq0.05$ at redshift $z\sim0.8$. Thus, foreground contamination can present a significant challenge to the clustering-redshifts method using two-point correlators involving 21 cm.

  \subsection{Clustering redshifts with three-point functions}\label{ssec:clustz.3pt}
    The non-linear nature of gravitational collapse leads to a statistical coupling between long- and short-wavelength modes of the density field. This can be understood intuitively: the presence of a long-wavelength fluctuation on a small patch is equivalent to shifting the mean background density, which affects the amplitude and growth of inhomogeneities. Thus the variations in the local power spectrum are statistically coupled to the density field on large scales, leading to a non-vanishing squeezed-limit bispectrum \citep{bernardeau2002, chiang2014, chiang2017, chiang2018, deputter2018}. This is the key fact behind the idea of large-scale density reconstruction: by examining the statistical correlations of the small-scale density fluctuations, we can build an estimator of the large-scale overdensity. In the simplest formalism, using only information encoded in the bispectrum, this takes the form of a quadratic estimator in which the long-wavelength overdensity is reconstructed from a quadratic combination of small-scale modes \citep{1203.3639,1305.4642,1403.3411,1511.04680,1610.07062,li2020}. More optimally, this can be done via forward modelling \citep{schmittfull2017, modi2019, modi2021}, making use of virtually all $N$-point correlators.

    We could thus reconstruct the long-wavelength fluctuations lost to foreground removal in the HI field by using quadratic combinations of the small-scale HI fluctuations. The resulting large-scale modes could then be correlated with the projected galaxy overdensity, in what would effectively amount to a three-point correlator between two short-scale HI modes and one long-wavelength galaxy overdensity mode, to constrain the properties of the galaxy redshift distribution. Since the constraints imposed by photo-$z$ uncertainties and foreground contamination affect only the radial $\apar{k}$ modes, without restricting the transverse modes to any particular range, rather than phrasing our forecasts in terms of this quadratic reconstruction, we will quantify the total amount of information enclosed in the bispectrum.

    The equivalent of Eq. \ref{eq:FisherP} for the bispectra is
    \begin{equation}\label{eq:FisherB}
      F^B_{\alpha\beta}=\sum_{{\bs X}{\bs X}'}\frac{A}{4\pi}\frac{1}{6}\int_0^\infty {\rm d}k_\perp\,{\rm d}q_\perp\,{\rm d}p_\perp\,\frac{k_\perp q_\perp p_\perp}{\pi^2A_T}\,\partial_\alpha {\cal B}^{XYZ}(k_\perp,p_\perp,q_\perp)\,\partial_\beta {\cal B}^{X'Y'Z'}(k_\perp,p_\perp,q_\perp)\,{\cal I}^{XX'}(k_\perp)\,{\cal I}^{YY'}(q_\perp)\,{\cal I}^{ZZ'}(p_\perp),
    \end{equation}
    where ${\cal B}^{XYZ}$ is the bispectrum between three 2D fields, $A_T=\frac{1}{2}\sqrt{2k_\perp^2q_\perp^2+2q_\perp^2p_\perp^2+2p_\perp^2k_\perp^2-k_\perp^4-q_\perp^4-p_\perp^4}$ is the area of a triangle with sides $(k_\perp,p_\perp,q_\perp)$, and ${\bs X} = \lbrace X, Y, Z\rbrace$.

    For the Gaussian selection function assumed here, as shown in Section \ref{app:math}, this reduces to the following expression for the variance of $\apar{r}^*$ and $\apar{\sigma}$:
    \begin{align}\label{eq:var_r_B}
      &{\rm Var}^{-1}_B(\apar{r}^*)=\frac{1}{6}\frac{A}{4\pi}\int {\rm d}k_\perp {\rm d}p_\perp {\rm d}q_\perp \frac{k_\perp p_\perp q_\perp}{\pi^2A_T}\,{\cal I}^{TT}(k_\perp)\int\frac{{\rm d}\apar{p}}{2\pi}\frac{{\rm d}\apar{k}}{2\pi}\apar{k}^2{\rm e}^{-\apar{k}^2\apar{\sigma}^2}\frac{B^2_{ghh}(\apar{k},\apar{p},-\apar{k}-\apar{p};k_\perp,p_\perp,q_\perp)}{P_{hh}(\apar{p},q_\perp)P_{hh}(-\apar{k}-\apar{p},p_\perp)},\\\label{eq:var_s_B}
      &{\rm Var}^{-1}_B(\apar{\sigma}^*)=\frac{1}{6}\frac{A}{4\pi}\int {\rm d}k_\perp {\rm d}p_\perp {\rm d}q_\perp \frac{k_\perp p_\perp q_\perp}{\pi^2A_T}\,{\cal I}^{TT}(k_\perp)\int\frac{{\rm d}\apar{p}}{2\pi}\frac{{\rm d}\apar{k}}{2\pi}\apar{\sigma}^2\apar{k}^4{\rm e}^{-\apar{k}^2\apar{\sigma}^2}\frac{B^2_{ghh}(\apar{k},\apar{p},-\apar{k}-\apar{p};k_\perp,p_\perp,q_\perp)}{P_{hh}(\apar{p},q_\perp)P_{hh}(-\apar{k}-\apar{p},p_\perp)},
    \end{align}
    assuming ${\rm Cov}(\apar{r}^*,\apar{\sigma})=0$. Here $B_{ghh}(\apar{k},\apar{q},\apar{p};k_\perp, q_\perp, p_\perp)$ is the bispectrum between the 3D galaxy overdensity $\delta_g({\bs k})$ and the HI field at two different wavenumbers ${\bs q}$ and ${\bs p}$.

    These equations demonstrate the ability of higher-order correlators to bypass the loss of radial modes to foregrounds affecting the 2-point function. Although, as in the case of the power spectrum, photometric uncertainties limit the range of usable small-scale radial modes of the galaxy overdensity through the exponential damping term, this does not limit the range of small-scale radial HI modes: $\apar{p}$ (and therefore $|-\apar{k}-\apar{p}|$) can take arbitrarily large values, deep into the region unaffected by foreground contamination. In other words, power spectra require $k_\parallel$ to be the same for the HI and galaxy overdensity fields, and the lack of overlap between them due to photometric redshift uncertainties and foreground mode loss suppresses the cross-correlation significantly. In turn, bispectra only force the radial mode of the galaxy field to be $\apar{k}^g=-\apar{k}^{h1}-\apar{k}^{h2}$, where $\apar{k}^{h1,2}$ are the radial wavenumbers for the two HI legs. While only small $\apar{k}^g$ survive due to photo-$z$ uncertainties, the HI modes can be arbitrarily large (thus avoiding the foreground-dominated regime) as long as they satisfy the previous equality.

\section{Survey modelling}\label{sec:surveys}
\subsection{Power spectra and bispectra of biased tracers}
  For generic biased tracers $x, y$ and $z$ of the dark matter field, the three-dimensional power spectrum and bispectrum are given by
  \begin{equation}
    P_{xy}(k) = {\cal K}_{xy}\, P_{\rm nl}(k) + N_{xy}\, \delta_{xy},\label{eq:generic_power}
  \end{equation}and
  \begin{align}
    B_{x y z}(\bs{k}_1, \bs{k}_2, \bs{k}_3) = {\cal K}_{xyz} B(\bs{k}_1, \bs{k}_2, \bs{k}_3) &+ {\cal J}_{xyz}^{112} P(k_1) P(k_2)+ {\cal J}_{xyz}^{121} P(k_1) P(k_3)+ {\cal J}_{xyz}^{211} P(k_2) P(k_3),\label{eq:generic_bisp}
  \end{align}
  respectively. In the above equations we are omitting the redshift dependence, which is implied in every quantity appearing on both correlators. We introduced the kernels ${\cal K}_{xy} \equiv {\cal K}_x\, {\cal K}_y$ and ${\cal K}_{xyz} \equiv {\cal K}_x\, {\cal K}_y\, {\cal K}_z$ to include particularities of each tracer, for example: ${\cal K}_{g} = b_{g,1}$ gives the linear bias for a particular galaxy sample, and ${\cal K}_h = b_{h,1}\bar{T}_h$ includes the mean temperature of the 21-cm signal. In Eq. \ref{eq:generic_power}, $N_{xy}$ is the power-spectrum noise which we neglect in the case $x\neq y$, as already mentioned in Section \ref{ssec:clustz.2pt}. 
  
  One may argue that, because the HI is tracing galaxies, there is the possibility of overlap between the photometric sample and the galaxies traced by the intensity mapping, which could lead to a correlated shot noise. However, even if this is the case, the Fisher matrix formalism that we base ourselves in is free from such contributions, since the noise in the signal will not depend on the parameters under consideration (i.e. it is independent of the photometric redshifts). Finally, although the ${\cal I}^{TT}$ contribution in the inverse covariance weighting of Eqs. \ref{eq:FisherP} and \ref{eq:FisherB} might depend on some correlated noise coming through the cross-correlation $P_{gh}$ (see Eq. \ref{eq:ITT_main}), only scales $k\lesssim 0.1\,\kunit$ will be affected, where the impact of stochastic noise should be unimportant. At these scales, the most relevant effect comes from foreground suppression, which overcomes other possible sources of noise.
  
  In Eq. \ref{eq:generic_bisp}, $B(\bs{k}_1, \bs{k}_2, \bs{k}_3)$ is the matter bispectrum induced by non-linear gravitational dynamics of the dark matter field. We do not include any noise component in $B_{ghh}$ (although shot noise and instrumental noise are fully taken into account in the Fisher matrix calculation) for two reasons: firstly, any noise contribution would not depend on the Fisher matrix parameters, and thus drops out when taking derivatives with respect to them. Moreover, the autocorrelation of galaxies in the bispectrum will not be considered, and thus shot-noise contributions from galaxies are irrelevant. The 21-cm instrumental noise is also irrelevant at the 3-point function level \citep{cunnington2021}. Finally, ${\cal J}_{xyz}^{ijk} \equiv {\cal J}_{x}^{i}{\cal J}_{y}^{j}{\cal J}_{z}^{k}$ accounts for the second-order bias $b_{x,2}$ contributions induced by the non-linear clustering. As an example: ${\cal J}_{ghh}^{112} \equiv b_{g,1}b_{h,1}b_{h,2}\,\bar{T}_h^2$.

  In the remaining of this work, we will adopt the fitting formula proposed in \citet{gilmarin2012}, based on $\Lambda$CDM dark-matter-only $N$-body simulations:
  \begin{equation}\label{eq:bispectrumFeff}
    B(\bs{k}_1, \bs{k}_2, \bs{k}_3) = 2F_2^{\rm eff}(\bs{k}_1, \bs{k}_2)P_{\rm nl}(k_1)P_{\rm nl}(k_2) + \text{ perms},
  \end{equation}with
  \begin{equation}
    F_2^{\rm eff}(\bs{k}_i,\bs{k}_j) = \frac{5}{7}\tilde{a}(k_i)\tilde{a}(k_j) + \frac{1}{2}\cos(\theta_{ij})\left(\frac{k_i}{k_j} + \frac{k_j}{k_i}\right) \tilde{b}(k_i)\tilde{b}(k_j) + \frac{2}{7} \cos^2(\theta_{ij}) \tilde{c}(k_i)\tilde{c}(k_j)
  \end{equation}recovering the tree-level result in the limit of $\tilde{a},\tilde{b},\tilde{c}\rightarrow 1$ (valid for large scales)\footnote{For more details on these functions, we refer the reader to \citet{gilmarin2012}.} and replacing the non-linear matter power spectrum $P_{\rm nl}(\bs{k})$ by the linear one. We also assume a flat $\Lambda$CDM model with $h = 0.674, n_s = 0.965, \sigma_8 = 0.811, \Omega_m = 0.315$, and $\Omega_b = 0.049$ \citep{planck2018}. Finally, non-linearities in the bias will be modelled via the phenomenological fit proposed by \citet{lazeyras2016}, which relates $b_2$ with the linear bias $b_1$ for dark matter haloes in a flat $\Lambda$CDM universe as
  \begin{equation}
    b_2(b_1) = 0.412 - 2.143 \,b_1 + 0.929 \,b_1^2+ 0.008 \,b_1^3. \label{eq:b2}
  \end{equation}
  Note that, although we shall assume that Eq. \ref{eq:b2} holds exactly for both galaxies and neutral hydrogen, its applicability in these cases is uncertain. This is particularly relevant when dealing with galaxies whose formation depends upon several astrophysical processes that may induce a large scatter in the bias relation \citep{zennaro2021}. This, however, allows us to roughly quantify the impact of non-linear bias on our results.

\subsection{Galaxy surveys: LSST}\label{ssec:lsst}
  Galaxy surveys are characterised by, among other things, the galaxy counts per unit redshift bin ${\rm d} z$ and solid angle ${\rm d}\Omega$, $n_g(z) = {\rm d}N/{\rm d} z/{\rm d}\Omega$. For this, we will assume the common parametrization \citep{ma2006}:
  \begin{equation}
     n_g(z) = A\, z^{\alpha} \, \exp\left[-\left(\frac{z}{z_0}\right)^{\beta}\right].\label{eq:true_dist}
  \end{equation}
  In our case, the redshift distribution sets the mean number density of galaxies at a given redshift, which determines the shot noise level in Eq. \ref{eq:generic_power},
  \begin{equation}Ŋ
      N_{gg} = \frac{1}{\bar{n}_g},
  \end{equation}
  after converting Eq. \ref{eq:true_dist} to the appropriate units.

  Our discussion will focus on the Vera Rubin Observatory's Legacy Survey of Space and Time (LSST). In this case, the redshift distribution is well approximated by the parameters $(z_0,\alpha,\beta)=(0.28,2,0.9)$. The linear bias will be modelled as \citep{1809.01669,ballardini2019,1912.08209}
  \begin{equation}
    b_{g,1}(z) = 0.95\, D^{-1}(z),\label{eq:b1_gal}
  \end{equation}
  where $D(z)$ is the growth factor normalised to $D(z=0)=1$. The second-order bias is then determined through Eq. \ref{eq:b2}.

  As stated in Section \ref{sec:clustz} (Eq. \ref{eq:phi_gaussian}), we will assume Gaussian redshift bins characterised by a mean redshift $\bar{z}$ and standard deviation $\sigma_z$, which can be related to the mean comoving distance and its scatter via $\apar{r}^*=\chi(\bar{z})$, $\apar{\sigma}=c\sigma_z/H(z)$, where $H(z)$ is the expansion rate and $\chi(z)$ is the radial comoving distance to redshift $z$. We will characterise the evolution of the standard deviation as $\sigma_z=\sigma_{z,0}(1+z)$, and we will consider values for $\sigma_{z,0}$ in the range $[0.02, 0.05]$, with a fiducial value $\sigma_{z,0}=0.03$. We will explore redshifts $z<2.5$, and we will fix the survey area to $\Omega_{\rm sky}=13{,}800\,{\rm deg}^2$.

\subsection{21-cm intensity mapping}
    \subsubsection{The cosmological HI signal}
        Intensity mapping of the HI signal maps the fluctuations $\delta T_h$ of the brightness temperature $T_h$ around its homogeneous component $\bar{T}_h$. Assuming that HI is a biased tracer of the dark matter field, the temperature fluctuations can be written as \citep{astro-ph/0411342}:
        \begin{equation}
            \delta T_h(\bs{x}) \equiv \bar{T}_h(z) \delta_h(\bs{x}) \equiv \bar{T}_h(z) \left[b_{h,1}(z)\delta(\bs{x}) + \frac{b_{h,2}(z)}{2} \delta^2(\bs{x})\right].
        \end{equation}
        The linear biasing scheme $b_{h,1}(z) \equiv b_h(z)$ and the mean brightness temperature $\bar{T}_h$ are taken from \citet{santos2017,ballardini2019}:
        \begin{equation}
            b_h(z) = 1.307\left(0.66655 + 0.17765\,z + 0.050223\,z^2\right),\label{eq:b1_hi}
        \end{equation}
        and
        \begin{equation}
            \bar{T}_h(z) = (0.055919 + 0.23242\, z - 0.024136\,z^2) \,\,[{\rm mK}]
        \end{equation}
        The second-order bias $b_{h,2}$ is obtained from Eq. \ref{eq:b2} after fixing $b_{h,1}$ to Eq. \ref{eq:b1_hi}.
    
    \subsubsection{Foreground contamination}
        Foregrounds originated from astrophysical emissions within and outside our Galaxy have a major impact on the observed temperature fluctuations $\delta T_h^{\rm obs}$. The cleaning of contaminated modes removes the cosmological signal on large radial scales, where it is confused with the spectrally smooth foregrounds \citep{petrovic2011,wolz2013,alonso2014,2019MNRAS.488.5452C,spinelli2021}. This effect can be modelled by introducing a damping function to the observed temperature fluctuation field:
        \begin{equation}
            \delta T_h^{\rm obs}(\bs{k}) = {\cal S}_{\rm FG}(\apar{k})\, \delta T_h(\bs{k}).\label{eq:deltaTh_onlyFG}
        \end{equation}
        
\begin{figure}
            \centering
            \includegraphics[width=\columnwidth]{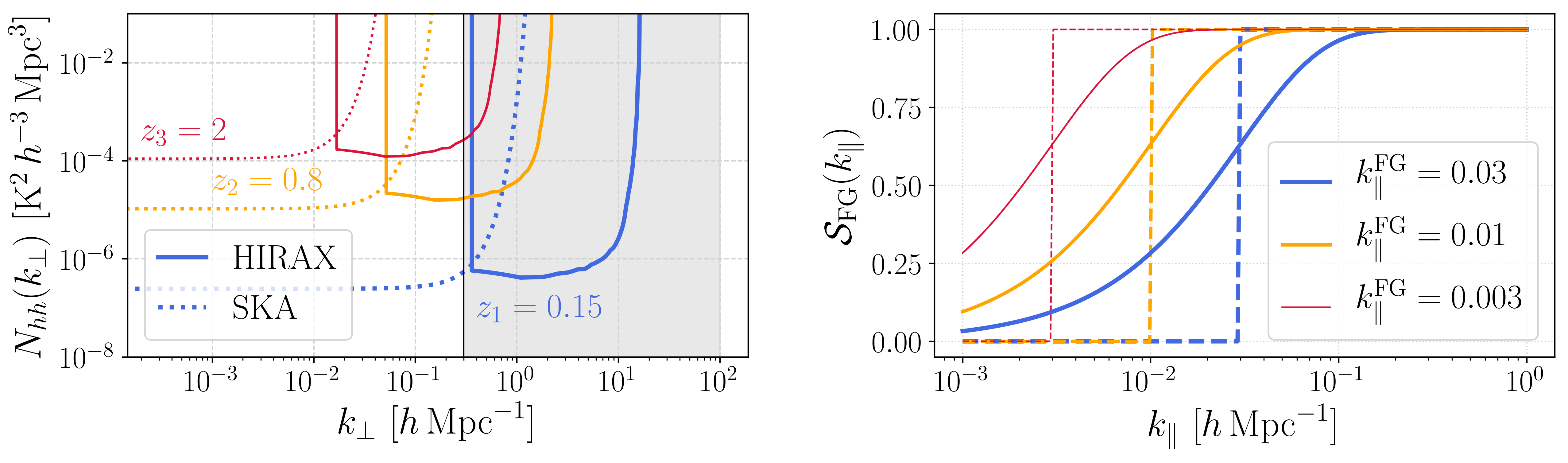}
            \caption{(\textit{Left}): HIRAX noise power spectrum $N_{hh}(k_\perp)$ (\textit{solid}), as described by Eq. \ref{eq:HIRAX_noise}, and the SKA noise (Eq. \ref{eq:SKA_noise}) including beam smoothing (i.e. $N_{hh}\,{\cal S}^{-2}_{\rm b}(k_\perp)$) (\textit{dotted}). The beam depicted is for SKA's single-dish mode, with fixed dish diameter $D_{\rm dish} = 15$ m. The shaded region corresponds to the modes eliminated by our fiducial  $k_{\rm max} = 0.3\,\kunit$ cutoff, but we stress that the $z_1=0.15$ will not be covered by HIRAX. Different colours correspond to different redshifts. (\textit{Right}): Foreground damping function, Eq. \ref{eq:FG_damping}. The functions ${\cal S}_{\rm FG}^{\rm HS}$ (\textit{dashed}) and ${\cal S}_{\rm FG}^{\rm SM}$ (\textit{solid}) are shown for different foreground damping scales $\apar{k}^{\rm FG}$, with $\apar{k}^{\rm min} = \apar{k}^{\rm FG}$ in the Heaviside model.}
            \label{fig:beam_FG}
        \end{figure}

        We will consider two cases based on the functions proposed in \citet{cunnington2020, cunnington2021}: \textbf{(1)} a conservative Heaviside (HS) model, in which all modes below a given scale are completely suppressed, and \textbf{(2)} a smooth (SM) model where large  radial modes are exponentially suppressed:
        \begin{equation}
            {\cal S}_{\rm FG}^{\rm HS}(\apar{k}) =  \begin{cases} \,\,\, 0, & \apar{k} < \kFG \\ \,\,\, 1, & \apar{k} > \kFG\end{cases},\hspace{12pt} \text{ and } \hspace{12pt}{\cal S}_{\rm FG}^{\rm SM}(\apar{k}) = 1-\exp\left[-\frac{\apar{k}}{\kFG}\right]. \label{eq:FG_damping}
        \end{equation}
        Above, $\kFG$ is the characteristic damping scale below which modes are suppressed after foreground removal. $\kFG=0$ represents the limit of perfect cleaning. The HS model is the most conservative case, assuming that no information can be retrieved below $\kFG$. In practice, some the cosmological signal will not be absolutely suppressed on large scales, and the SM model aims to represent this. However, the level of suppression will depend on multiple factors, including foreground complexity, instrumental effects (e.g. polarisation leakage or frequency-dependent beams), and choice of cleaning method. It is not obvious that the corresponding transfer function can be sufficiently well characterised in the foreground-dominated regime, and therefore we will use the HS model as the fiducial scenario in our analysis. Both choices are shown in the right panel of Fig. \ref{fig:beam_FG} for various values of $\kFG$.

        Another constraint on the modes accessible to the IM survey related to foregrounds comes in the form of the so-called ``wedge effect''. This is an issue associated with the frequency dependence of the transverse modes induced by the chromatic response of the experiment \citep{seo2016}. For the case of interferometry, we consider the case where all modes \citep[see][and references therein]{alonso2017}
        \begin{equation}
        	\apar{k}<\apar{k}^{\rm hor} \equiv \frac{\chi(z)\,H(z)}{c\,(1+z)}k_\perp\label{eq:horizon_wedge}
        \end{equation}become inaccessible by the survey, defining the ``horizon wedge''. It appears from a phase-delay modulation ${\rm e}^{2\pi\,i\,\bs{u}\cdot\bs{\theta}}$ induced by the fringe pattern of the telescope. Here, $\bs{u}=\bs{d}\,\nu/c$ is the baseline vector, with $\bs{d}$ the ground distribution vector of antennas, and $\bs{\theta}$ is the 2-dimensional source location \citep[e.g., see][]{choudhuri2021}. Because $2\pi\bs{u}$ sets the angular resolution, associated with the transverse modes $\aperp{k}$, when we cross-correlate the signal received by different antennas, the frequency structure gets shifted and a foreground, in principle smooth in frequency, acquires a chromatic response. Therefore, we can use the model shown in Eq. \ref{eq:horizon_wedge} to represent the radial modes $\apar{k}^{\rm hor}$ we lose due to this modulation, which depends on the perpendicular scales $\aperp{k}$. Nevertheless, note that various systematics could lead to a loss of modes beyond the wedge as defined in Eq. \ref{eq:horizon_wedge}, and a thorough study of these will be necessary to fully quantify the feasibility of the method proposed here in practice.
        
        The second modulation is due to the telescope's primary beam, affecting both the interferometer and single-dish observations: since the telescope's beam has a frequency dependence, if we observe a foreground at a fixed sky position $\bs{\theta}$, its spectrum can be more smoothed if it falls near the side lobes of the beam, or more enhanced by falling near the central lobe's peak. This effect adds chromaticity to a smooth foreground. For this ``primary-beam wedge'', the excluded modes will be modelled as 
        \begin{equation}
        	\apar{k} < \apar{k}^{\rm pb} \equiv \sin\left(\frac{\theta_{\rm FWHM}}{2}\right)\apar{k}^{\rm hor},
        \end{equation}where $\theta_{\rm FWHM}$ is the beam width, in radians, defined at its full-width-half-maximum (FWHM): $\theta_{\rm FWHM}(z) = 0.21 (1+z) \,D_{\rm dish}^{-1}$. The primary beam wedge is therefore a less restrictive version of the horizon wedge.

    \subsubsection{Single-dish experiments: SKA}
        We consider SKA1-MID in the single-dish mode\footnote{Note that it is also possible to carry out intensity mapping observations with SKA in interferometer mode. Unfortunately, the SKA baseline distribution is not optimised for cosmology, and most scales used here ($k_\perp<0.3\,\kunit$) are lost in this case. This could be revisited if the approach described here turned out to be applicable with sufficient robustness on small scales.}. This experiment consists of $N_{\rm dishes} = 133$ dishes with $D_{\rm dish} = 15$ m in diameter. While it is required Bands 1 and 2 to fully observe the redshift range $0 < z < 3$ \citep{santos2015}, for simplicity we shall consider Band 1 specifications only: it has a sky coverage of $\Omega_{\rm tot} = 20000$ deg$^2$, frequency resolution of $\Delta \nu = 15.2$ kHz, and integration time of $t_{\rm tot} \approx 10000$ hours \citep{bacon2020}.
        
        The finite resolution of the telescope's primary beam removes structure in the temperature fluctuations on small angular scales. The impact of this smoothing can be taken into account by modifying Eq. \ref{eq:deltaTh_onlyFG} as:
        \begin{equation}
            \delta T_h^{\rm obs}(\apar{k},\aperp{k}) = {\cal S}_{\rm b}(k_\perp)\, {\cal S}_{\rm FG}(\apar{k})\, \delta T_h(\apar{k},\aperp{k}),\label{eq:deltaTh}
        \end{equation}
        where ${\cal S}_{\rm b}(k_\perp)$ is the beam window function. We will assume a Gaussian beam of the form ${\cal S}_{\rm b}(k_\perp)=\exp(-(k_\perp\sigma_\perp)^2/2)$,
        where $\sigma_\perp$ is related to $\theta_{\rm FWHM}$ via
        \begin{equation}
          \sigma_\perp = \chi(z)\frac{\theta_{\rm FWHM}}{2\sqrt{2\log2}}.
        \end{equation}
        
        For the IM, shot noise is not a relevant contribution. The noise in Eq. \ref{eq:generic_power} comes from thermal fluctuations in the instrument, acting as a source of white noise. For the single-dish mode, it is described by a Gaussian field with variance \citep{pourtsidou2017}
        \begin{equation}
            \sigma_{\mathrm{pix}}^2 = T_{\mathrm{sys}}^2 \frac{1}{\Delta\nu\, t_{\mathrm{tot}}} \frac{\Omega_{\mathrm{tot}}}{\Omega_{\mathrm{pix}}} \frac{1}{N_{\mathrm{dishes}} N_{\mathrm{beams}}},
        \end{equation}
        where $\Omega_{\rm pix} = 1.133\, \theta^2_{\rm FWHM}$ for a Gaussian beam, $T_{\rm sys} = T_{\rm rx} + T_{\rm spill} + T_{\rm CMB} + T_{\rm gal}$ encodes the temperature contributions from the ground (spill-over $T_{\rm spill}\approx 3$ K), from the CMB ($T_{\rm CMB} \approx 2.7$ K), and from our Galaxy ($T_{\rm gal}$) and the system's noise ($T_{\rm rx}$) \citep{bacon2020}:
        \begin{equation}
            T_{\rm gal} = 25\, \left(\frac{408 \, {\rm MHz}}{\nu}\right)^{2.75}\,{\rm K},\hspace{12pt}\text{ and }\hspace{12pt} T_{\rm rx} = 15\,{\rm K} + 30\,{\rm K} \left(\frac{\nu}{\rm GHz} - 0.75\right)^2.
        \end{equation}
        In the remaining of this work, we consider $N_{\rm beams} = 1$.
        
        Therefore, the noise in the autocorrelation of HI maps is
        \begin{equation}
            N_{hh} = \sigma_{\rm pix}^2\, V_{\rm pix},\label{eq:SKA_noise}
        \end{equation}where
        \begin{equation}
            V_{\rm pix} = \Omega_{\rm pix} \int_{z_{\rm min}}^{z_{\rm max}} {
            \rm d}z\frac{c\,\chi^2(z)}{H(z)}
        \end{equation}is the comoving volume of a ``pixel''. The dotted curves in the left panel of Fig. \ref{fig:beam_FG} shows the SKA noise, Eq. \ref{eq:SKA_noise}, and the beam smoothing effect, i.e. $N_{hh}(k_\perp) \equiv N_{hh}\,{\cal S}_{\rm b}(k_\perp)$.

\subsubsection{Interferometers: HIRAX}\label{sssec:hirax}

The Hydrogen Intensity mapping and Real-time Analysis eXperiment \citep[HIRAX,][]{2016SPIE.9906E..5XN,crichton2021} is a close-packed array of radio antennas in a square $32\times32$ configuration. Each of the 6-m dishes will take measurements of the radio sky over $15{,}000$ deg$^2$ in the frequency band $400-800$ MHz, corresponding to the 21-cm line in the range of redshifts $0.8\lesssim z \lesssim 2.5$. Like SKA, HIRAX is being deployed in the South African Karoo site, which provides an ideal location in terms of RFI contamination, as well as area overlap with LSST. Besides the standard 21-cm cosmological science cases enabled by its large collecting area and field-of-view, the resulting mapping speed will allow HIRAX to produce large catalogues of localised Fast Radio Bursts \citep{2019arXiv190507132W}. Here we explore its ability to provide calibrated constraints on the redshift distribution of photometric galaxy samples through the clustering redshifts approach using the power spectrum and bispectrum in the presence of realistic foreground contamination.

\begingroup
\allowdisplaybreaks
We model the noise properties of HIRAX following the model described in Appendix C of \citet{alonso2017}:
\begin{equation}
  N_{hh}(k_\perp)=\frac{4\pi f_{\rm sky}\,\chi^2(z)\,(1+z)\,T_{\rm sys}^2\,\theta_{\rm FWHM}^2}{H(z)\,t_{\rm tot}\,\lambda_{21}(z) \, N_{d}({\bs d}={\bs k}_\perp\,\chi(z)\,\lambda_{21}(z)/2\pi)},\label{eq:HIRAX_noise}
\end{equation}
where $N_d({\bs d})$ is the number density of baselines (i.e. separations between pairs of antennas) with distance ${\bs d}$, $\lambda_{21}(z)$ is the observed 21-cm line wavelength and we assume $T_{\rm sys} = 50 \, {\rm K} + 60\,(\nu_{21}(z)/300\,{\rm MHz})^{-2.5}\, {\rm K}$. For simplicity we assume an isotropic density of baselines computed from the distribution corresponding to a $32 \times 32$ tightly-packed interferometer (i.e. assuming a 6-m separation between antennas). The solid curves in the left panel of Fig. \ref{fig:beam_FG} shows the HIRAX noise, Eq. \ref{eq:HIRAX_noise}, as a function of the perpendicular modes. 
\endgroup

The advantage of HIRAX over SKA is its ability to cover significantly smaller angular scales, in the range of interest for cosmological studies. Whereas the single-dish SKA beam smooths out comoving scales above $k^{\rm SKA}_\perp\simeq 0.045\,\kunit$ at $z=1$ ($\theta_{\rm FWHM}=1.6^\circ$), HIRAX can recover the 21-cm fluctuations out to $k^{\rm HIRAX}_\perp\simeq1\,\kunit$ at the same redshift. The drawback is that interferometers are not able to resolve scales larger than that determined by the minimum available baseline (corresponding to $k_\perp\sim0.037\,\kunit$ at $z=1$ for HIRAX). However, since the number of available modes grows steeply with the maximum $k_\perp$, as does the amplitude of the bispectrum as we approach non-linear scales, we can expect a significant improvement over SKA when using HIRAX, particularly at high redshifts.

\begin{figure}
  \centering
  \includegraphics[width=0.6\textwidth]{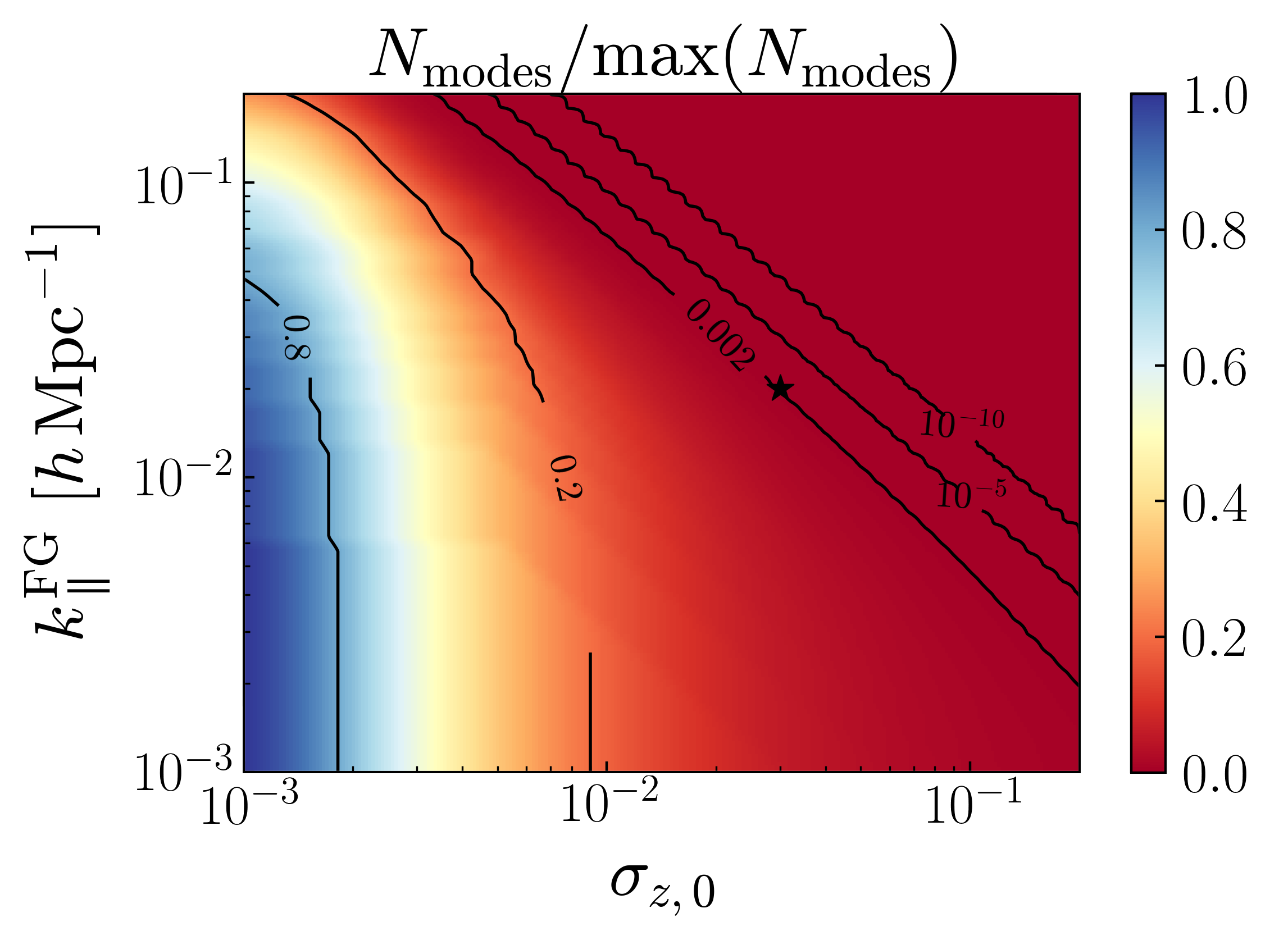}
  \caption{Fractional number of modes available to the galaxy-HI cross-correlation for a HIRAX-like experiment as a function of the photometric redshift error ($x$-axis) and the foreground cutoff scale ($y$-axis). The black star marks the fiducial case explored here: $(\sigma_{z,0},\kFG)= (0.03, 0.02\,\,\kunit)$.}\label{fig:nmodes}
\end{figure}

\section{Results}\label{sec:results}

\subsection{Mode-loss and mode reconstruction}\label{ssec:simplified_results}

In order to determine the regime in which long-wavelength reconstruction via the bispectrum is able to outperform the standard cross-correlation redshifts approach, we start by considering an idealised galaxy sample centred at mean redshift $z=0.8$, for which Eq. \ref{eq:b1_gal} corresponds to a constant linear bias $b\approx 1.4$, and characterised by a Gaussian window function. We will study the cross-correlation of this sample with a HIRAX-like intensity mapping experiment (in terms of angular resolution and noise), varying the two characteristic scales that determine the overlap in $k$ space between both datasets: the width of the Gaussian kernel $\sigma_{z,0}$, the foreground cutoff scale $\kFG$, and the smallest scale used in the analysis $k_{\rm max}$ (for both radial and perpendicular directions). For completeness, in Appendix \ref{ap:extra} we present the same results, but considering an SKA-like survey.

Figure \ref{fig:nmodes} shows the number of modes available to the cross-correlation between both samples, given by
\begin{equation}
  N_{\rm modes} = \frac{V}{2\pi^2}\int_0^{k_{\rm max}}{\rm d}\apar{k}\int_0^{k_{\rm max}}{\rm d}k_\perp\,\,k_\perp\, {\rm e}^{-\apar{k}^2\apar{\sigma}^2}\frac{P^2_{gh}(\apar{k},\aperp{k})}{P_{gg}(\apar{k},\aperp{k})P_{hh}(\apar{k},\aperp{k})},
\end{equation}
where $V$ is the survey volume and we use $k_{\rm max}=0.3\,\kunit$. Both $P_{gh}$ and $P_{hh}$ incorporate the foreground window function, and the angular beam, and the figure shows $N_{\rm modes}$ as a function of $\sigma_{z,0}$ and $\kFG$ (for the Heaviside cutoff) and as a fraction of the total number of available modes (when $\kFG=0$, $\sigma_{z,0}=0$). For realistic values of these characteristic scales ($\kFG\sim0.01\,\kunit$, $\sigma_{z,0}\gtrsim0.01$) there is a sharp decrease in the fractional number of modes that can be probed through the galaxy-HI cross-power spectrum, reaching $\sim0.5\%$ for the fiducial case used in this paper: $\sigma_{z,0}=0.03$ (an optimistic estimate of the photometric redshift accuracy achievable by LSST), and $\kFG=0.02\,\kunit$ \citep{2015PhRvD..91h3514S}. This fiducial case is shown as a black star in the figure.

\begin{figure}
  \centering
  \includegraphics[width=\textwidth]{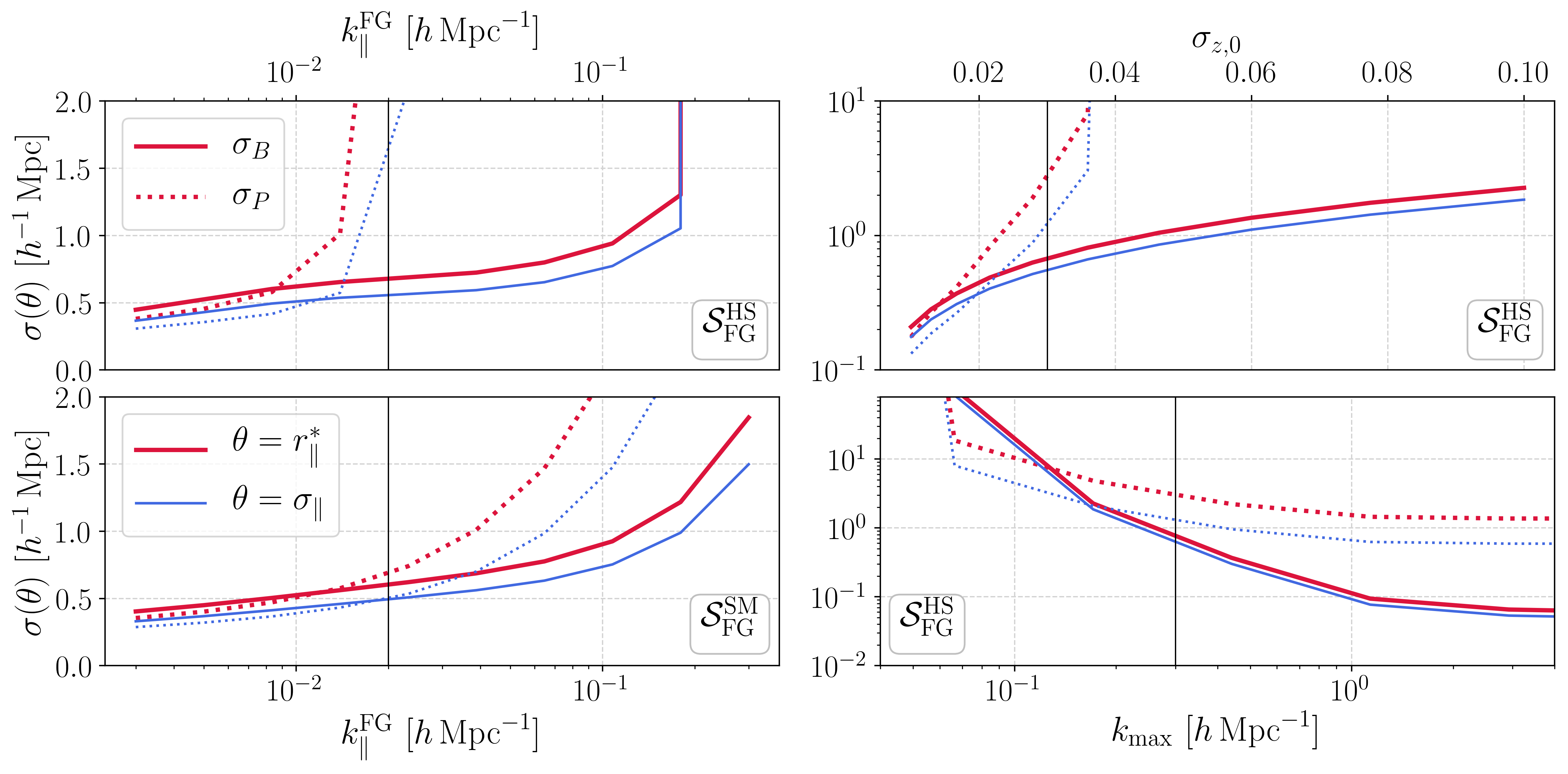}
  \caption{Power-spectrum (\textit{dotted}) and bispectrum (\textit{solid}) constraints obtained by cross-correlating a HIRAX-like (interferometer) survey and a LSST-like, on both Gaussian photo-$z$ parameters, $\theta = \apar{r}^*$ (\textit{thick red}) and $\theta = \apar{\sigma}$ (\textit{thin blue}), as a function of foreground cutoff/damping scale $\kFG$ ({\it left, top and bottom}), photo-$z$ scatter $\sigma_{z,0}$ ({\it right, top}), and maximum wavenumber $k_{\rm max}$ ({\it right, bottom}). In the top-left panel, foregrounds are treated via the Heaviside model, ${\cal S}_{\rm FG}^{\rm HS}(\apar{k})$ in Eq. \ref{eq:FG_damping}, while the bottom panel shows the same constraints for the smooth exponential damping, ${\cal S}_{\rm FG}^{\rm SM}(\apar{k})$. The plots on the right consider only the ${\cal S}_{\rm FG}^{\rm HS}(\apar{k})$ modelling. A single redshift bin centred at $z=0.8$ is assumed. Linear bias and number density for the photometric galaxies are, respectively, $b \approx 1.4$ and $\bar{n} \approx 0.004$. We neglect second-order bias contributions and wedge effects in this plot. The vertical lines in each panel show the fiducial values of these scales used in our analysis.}
  \label{fig:HIRAX_idealised}
\end{figure}

To quantify to what extent the $ghh$ bispectrum is able to overcome this loss of modes, we study the dependence of the errors on the photo-$z$ parameters $(\apar{r}^*,\apar{\sigma})$ as a function of the characteristic scales $(\sigma_{z,0}, \kFG, k_{\rm max})$. Figure \ref{fig:HIRAX_idealised} shows the standard deviation of $\apar{r}^*$ (thick and red) and $\apar{\sigma}$ (thin and blue) for the power spectrum (dotted curves) and bispectrum (solid curves) in four different cases. In each case, the scales not varied are fixed to the fiducial values $\sigma_{z,0}=0.03$, $\kFG=0.02\,\kunit$, and $k_{\rm max}=0.3\,\kunit$. The top-right panel shows the dependence on the photo-$z$ width. The errors on the photo-$z$ parameters from the power spectrum grow rapidly, diverging around $\sigma_{z,0}\sim0.02$, corresponding to a comoving scale $r \sim 68\,h^{-1}{\rm Mpc}$ at $z=0.8$. In turn, the bispectrum errors show a significantly milder dependence on the photo-$z$ scale, outperforming the power spectrum beyond $\sigma_{z,0}\sim0.02$.

The top-left panel of Fig. \ref{fig:HIRAX_idealised} shows the dependence of the errors in the photo-$z$ parameters on $\kFG$. Although the errors derived from the cross-power spectrum diverge for $\kFG\sim0.02\,\kunit$, roughly corresponding with the physical scale of the photo-$z$ scatter, the bispectrum constraints show a much milder growth down to the maximum scale $k_{\rm max}=0.3\,\kunit$. While this panel assumes a conservative binary foreground cutoff (i.e. foregrounds completely eliminate the signal below $\kFG$), the bottom-left panel of the figure shows the same result for the smooth exponential tapering off. In this case, the bispectrum constraints do not overtake the power spectrum until $\kFG\sim0.02\,\kunit$. This is due to the low noise of both HIRAX and LSST, which always allows us to make use of a sufficiently large number of modes, even if the signal has been suppressed by foreground removal. This, however, assumes that it is possible to model the post-cleaning signal accurately enough on scales $k_\perp\ll \kFG$, which is likely not the case in a realistic scenario. For this reason, we will assume the more conservative binary foreground cutoff model in the rest of the paper. The figure, however, shows that the quantitative results presented here are very sensitive to the assumptions made about the efficiency of the foreground removal procedure.

Finally, the bottom-right panel of Fig. \ref{fig:HIRAX_idealised} shows the dependence of the photo-$z$ parameter errors on the maximum wavenumber used in the analysis. We observe that, while the power spectrum constraints do not improve significantly beyond $k_{\rm max}\sim0.3\,\kunit$, the bispectrum errors improve steadily out to $k_{\rm max}\sim1\,\kunit$, where they saturate. This is mainly because, as before, nothing limits the smallest radial scale on which the HI fluctuations can be used in the bispectrum, other than the scale at which the 21-cm noise dominates. Besides that, although the cross-power spectrum is only significant out to the smallest transverse scale mapped by the 21-cm experiment, the triangular inequality allows us to use angular scales that are up to twice as small in the case of the bispectrum.

\subsection{Forecasts for Stage-IV surveys}\label{ssec:stageIV}

\begin{figure}
  \centering
  \includegraphics[width=\textwidth]{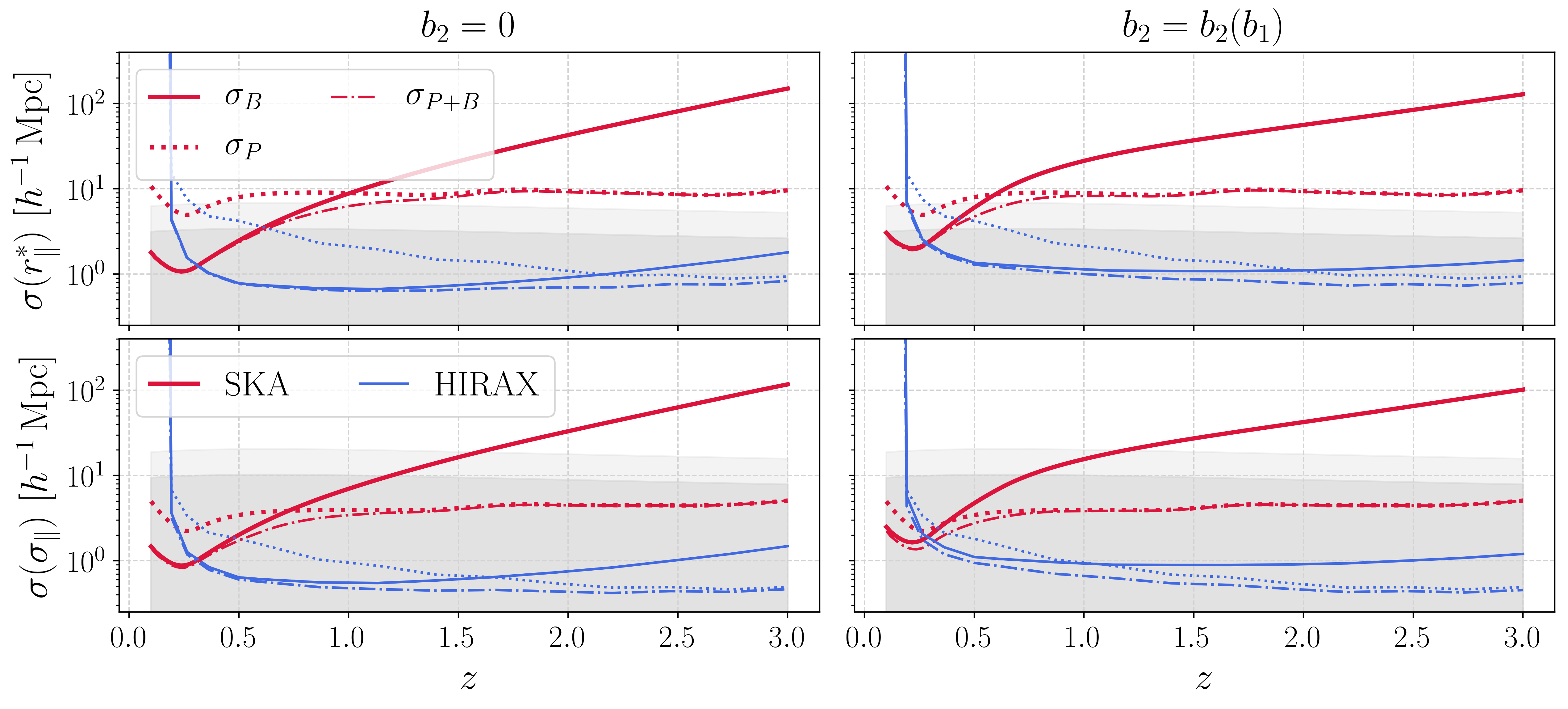}
  \caption{Constraints for the power spectrum (\textit{dotted}), bispectrum (\textit{solid}), and the combination of both statistics (\textit{dot dashed}) obtained with the HIRAX (\textit{thin blue}) and SKA (\textit{thick red}) IM surveys, in the absence of wedge effects. For the latter, we either neglect the second-order bias (\textit{left}), or include it as a function of the linear bias, following Eq. \ref{eq:b2} (\textit{right}). Light-shaded grey regions correspond to the year 1 (Y1) LSST requirements for weak lensing and LSS (3$\times$2-point) analyses, which are the most stringent ones, whereas darker regions correspond to these requirements for LSST Y10. Characteristic scales are fixed at the fiducial values $\sigma_{z,0} = 0.03,\apar{k}^{\rm FG} = 0.02\,\kunit$, and $k_{\rm max} = 0.3\,\kunit$. Note that, although we show results for the whole redshift range, HIRAX will only take measurements in the range $0.8<z<2.5$.}
  \label{fig:forecast_HIRAXSKA}
\end{figure}

Having gained some intuition about the performance of the clustering redshift method using power-spectra and bispectra as a function of photo-$z$ uncertainty and foreground mode-loss, we now proceed to present forecasts for the errors $\sigma(r^*_\parallel)$ and $\sigma(\sigma_\parallel)$ in the case of Stage-IV experiments. In particular, we will consider the cross-correlation of LSST data with SKA (single-dish mode) and HIRAX (interferometer), as a function of redshift. The radial distribution assumed for the LSST galaxies and their linear bias are described in Section \ref{ssec:lsst}. For the intensity mapping surveys, the foreground cutoff scale is fixed at $\apar{k}^{\rm FG} = 0.02\,\kunit$ \citep{2015PhRvD..91h3514S}, and we always assume $k_{\rm max} = 0.3\,\kunit$.

To assess the efficiency of different experiments to constrain the redshift distribution parameters, $\apar{\sigma}$ and $\apar{r}^*$, in Fig. \ref{fig:forecast_HIRAXSKA} we show the forecast results for both the SKA (thick and red) and HIRAX (thin and blue) surveys, in the wedge-free scenario, fixing the photo-$z$ scatter to $\sigma_{z,0} = 0.03$. The dark and light grey bands show the requirement for LSST in Year-1 (Y1) and Year-10 (Y10). No wedge effects were included for this figure. The top and bottom panels show the constraints on the bin's mean $\apar{r}^*$ and width $\apar{\sigma}$. The left and right panels show constraints considering only a linear bias, and including second-order bias terms respectively.

Let us focus on the $\apar{r}^*$ constraints (top panels), for which the LSST requirements are significantly tighter. As shown in the figure, due to the loss of radial modes SKA would not be able to achieve the LSST photo-$z$ calibration requirements using two-point correlations. However, utilising the information available in the 3-point function it would be possible to outperform the power-spectrum results out to $z\sim0.5$. At higher redshifts two factors conspire to degrade the bispectrum constraints achievable by SKA: the angular resolution decreases significantly, and the amplitude of the matter bispectrum decreases as the density field becomes more linear (see Appendix \ref{ap:snr} for a quantitative discussion). In Fig. \ref{fig:forecast_HIRAXSKA}, we also showcase the constraining power of a joint $2+3$-point analysis (dash-dotted curves), where the gains of each correlator are combined.

In the case of HIRAX, with access to smaller angular scales, the bispectrum outperforms the power spectrum out to $z\sim1.5-2$. The divergence around $z\sim 0.15$ for HIRAX is expected: we ignore all information on scales $k_\perp\gtrsim k_{\rm max} = 0.3\,\kunit$, which at this redshift fall below the minimum baseline of the experiment\footnote{Nevertheless, note that HIRAX will only target the redshift range $0.8\lesssim z \lesssim 2.5$.}. Note that the forecasts seem to indicate that, even if better constraints could be achieved with the bispectrum, HIRAX would be able to satisfy the LSST requirements even with the power spectrum. This statement, however, is heavily dependent on the assumed photo-$z$ width.

\begin{figure}
  \centering
  \includegraphics[width=\textwidth]{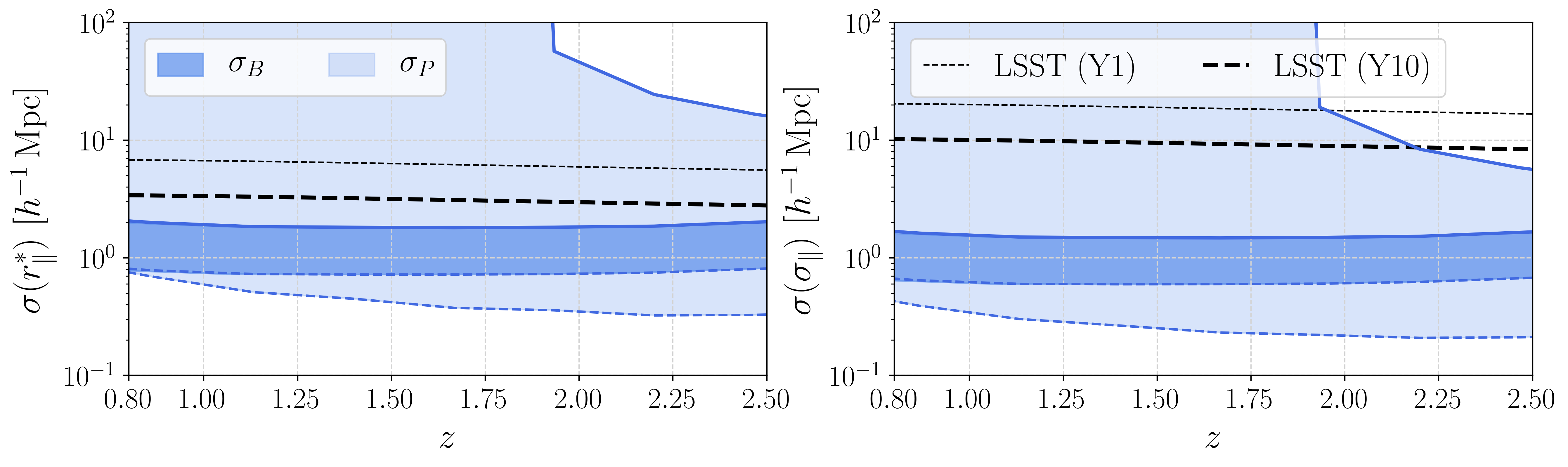}
  \caption{Constraints for the power spectrum (\textit{light}) and bispectrum (\textit{dark}) obtained with a HIRAX-like survey, in the absence of wedge effects, including the second-order bias according to Eq. \ref{eq:b2}. We show the regions of constraints extending from an optimistic photo-$z$ scatter ($\sigma_{z,0} = 0.02$, lower limit, \textit{blue-dashed} curves) up to a pessimistic scatter ($\sigma_{z,0} = 0.05$, upper limit, \textit{blue solid}), while keeping the other scales fixed at the fiducial values: $\apar{k}^{\rm FG} = 0.02\,\kunit$, and $k_{\rm max} = 0.3\,\kunit$. Black-dashed curves indicate the Y1 (\textit{thin}) and Y10 (\textit{thick}) requirements for the LSST's 3$\times$2-point analyses. }
  \label{fig:forecast_HIRAX}
\end{figure}

Figure \ref{fig:forecast_HIRAX} shows the constraints on the photo-$z$ parameters achievable by HIRAX for a photo-$z$ width $\sigma_{z,0}$ in the range $[0.02,0.05]$. Constraints are shown for the power spectrum (light blue) and bispectrum (dark blue). The power-spectrum constraints have a steep dependence on the photo-$z$ width, and achieving the LSST requirement would be impossible if $\sigma_{z,0}\gtrsim0.05$. Note that this is a perfectly realistic prospect. On the other hand, the bispectrum constraints show a much milder dependence on $\sigma_{z,0}$, and fall within the requirement for all $z$ and $\sigma_{z,0}$.

\begin{figure}
    \centering
    \includegraphics[width=\columnwidth]{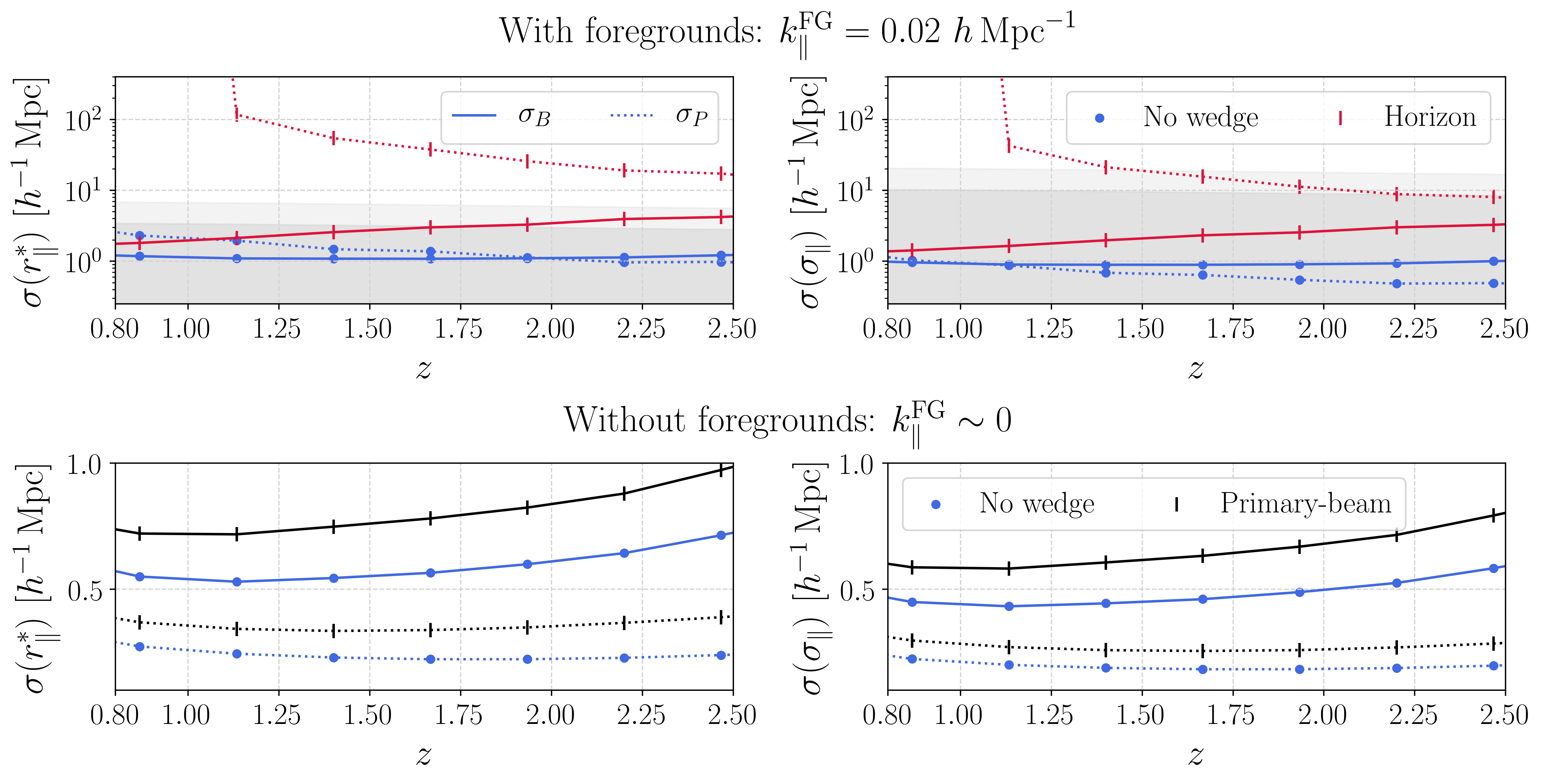}
    \caption{Constraints obtained from the cross-correlations with HIRAX for the power spectrum (\textit{dotted}) and bispectrum (\textit{solid}). The second-order bias correction is included as a function of the linear bias (Eq. \ref{eq:b2}). (\textit{Top}): In these plots we explore the effect of horizon wedge (\textit{red bars}) and the wedge-free case (\textit{circles}). Light-shaded regions correspond to the Y1 LSST requirements for the 3$\times$2-point analyses, whereas darker regions correspond to Y10. Characteristic scales are fixed at the fiducial values $\sigma_{z,0} = 0.03,\apar{k}^{\rm FG} = 0.02\,\kunit$, and $k_{\rm max} = 0.3\,\kunit$. (\textit{Bottom}): Constraints from HIRAX in the absence of foreground mode loss ($\apar{k}^{\rm FG} \sim 0$), and for $\sigma_{z,0} = 0.03$ and $k_{\rm max} = 0.3\,\kunit$. We took the primary-beam wedge (\textit{black bars}) as the only source of radial-mode loss. The wedge-free case (\textit{circles}) is shown for comparison.}
    \label{fig:forecast_HIRAXwedges}
\end{figure}

Let us now turn to the impact of the foreground wedge. For both the single-dish and interferometer experiments, the primary-beam wedge is irrelevant, since the modes affected by it are already suppressed by our choice of $\kFG=0.02\,\kunit$. The impact of the horizon wedge on HIRAX is shown in the top panels of Fig. \ref{fig:forecast_HIRAXwedges}. The presence of the horizon wedge has a strong impact on the power-spectrum constraints, which are degraded by more than one order of magnitude, falling outside the LSST requirement at all redshifts. The bispectrum constraints, on the other hand, are significantly less sensitive to the wedge, and the Y10 requirements are still achieved at $z\lesssim2.2$. 

For completeness, the bottom panels of Fig. \ref{fig:forecast_HIRAXwedges} present the forecast for HIRAX considering primary-beam wedge effects as the only source of foreground mode suppression (i.e. $\apar{k}^{\rm FG} \sim 0$), in which case the primary-beam wedge does have a relevant effect. In these panels, the constraints are all within the LSST's Y10 requirements: $\sigma(\apar{r}^*)\sim 3\,h^{-1}\,{\rm Mpc}$ and $\sigma(\apar{\sigma}) \sim 10\,h^{-1}\,{\rm Mpc}$. We do not show the horizon wedge for this foreground free scenario since the constraints are the same shown in the top panels of Fig. \ref{fig:forecast_HIRAXwedges}, meaning that, for the HIRAX IM survey, the horizon wedge is the leading cause of mode loss. 

Before finishing this section, it is worth turning back to the right panels in Fig. \ref{fig:forecast_HIRAXSKA}, showing the constraints accounting for quadratic bias contributions. Although we have shown that the bispectrum can outperform the two-point function for redshift distribution calibration in the presence of foregrounds, the quantitative value of the achieved constraints depends on the non-linear modelling of the galaxy and HI overdensities. Thus the final numbers presented here should only be interpreted as a rough order-of-magnitude estimation of the achievable constraints. In fact, assuming the optimistic value $\sigma_{z,0} = 0.03$, the power spectrum always provides better constraints on the photo-$z$ parameters for a cautious $k_{\rm max} \sim 0.15\,\kunit$, for both SKA and HIRAX (in the absence of a horizon wedge). But, as already discussed, the power spectrum for both surveys is unable to reach the required calibration constraints for $\sigma_{z,0} \sim 0.05$, or when a horizon wedge is present. In the extreme scenario, in which $\sigma_{z,0} \approx 0.05$ and there is a horizon wedge, we are unable to reach the desired constraints -- even with the bispectrum -- unless we go into the mildly non-linear regime ($k \sim 0.3\,\kunit$).

\section{Conclusions}\label{sec:conclusions}
The loss of long radial modes to foreground removals hampers the ability of cross-correlations between 21-cm intensity maps and photometric redshift surveys to calibrate the redshift distribution of the latter. In this paper we have explored the possibility of overcoming this issue through the use of higher-order correlations, focusing on the case of the bispectrum. The motivation for doing so is based on the idea of mode reconstruction: the statistical coupling between large- and small-scale modes makes it possible to infer the amplitude of a long-wavelength density fluctuation from the local variance of two small-scale modes.

With this in mind, we have quantified the information content of the bispectrum involving a map of the projected galaxy distribution and two HI maps, and propagated it into forecast constraints on redshift distribution parameters from the cross-correlation of LSST with HIRAX and SKA. In particular, we have used a Fisher matrix approach, focusing on two parameters of a simplified Gaussian radial selection function: its mean comoving distance $\apar{r}^*$ and standard deviation $\apar{\sigma}$.

Any constraints based on the two-point cross-correlation are severely hampered whenever the foreground cutoff scale $\kFG$ approaches or exceeds the scale associated with the redshift scatter $\sim1/\apar{\sigma}$. For typical photo-$z$ uncertainties ($\sigma_z\sim0.05$), this happens at $\kFG\gtrsim0.02\,\kunit$, comparable with the performance of usual foreground removal methods \citep{2015PhRvD..91h3514S}. In this regime, we have shown that the bispectrum is generally able to outperform power-spectrum constraints, and to calibrate the redshift distribution properties below the requirements of LSST.

This is true at low redshifts ($z\lesssim0.5$) for SKA in single-dish mode, although the constraints degrade significantly at higher redshifts due to the size of the SKA beam. In contrast, a HIRAX-like experiment in interferometer mode would be able to satisfy the LSST requirements throughout its whole redshift range ($0.8<z<2.5$) for redshift widths $\sigma_{z,0}\in[0.02,0.05]$ using the bispectrum, whereas this would be impossible with the two-point function. This result also holds in the presence of a horizon foreground wedge. Whereas the wedge has a devastating effect on the two-point function constraints, the corresponding mode loss is less severe for the bispectrum, which is still able to achieve the LSST requirements. The impact of the primary beam wedge is insignificant in all cases studied except when assuming a very optimistic $\kFG$.

Our fiducial forecasts used only scales $k<0.3\,\kunit$, where it should be possible to characterise the two- and three-point correlations with existing tools. These constraints can therefore be improved by including smaller scales, as long as the impact of non-linearities can be quantified. Nevertheless, we find that, given the expected sensitivity of SKA and HIRAX, the constraining power saturates at $k_{\rm max}\sim1\,\kunit$. As we showed, it is important to use scales slightly inside the non-linear regime ($k_{\rm max} \sim 0.3\,\kunit$), as the constraints can be severely degraded if only modes $k \sim 0.15\,\kunit$ are used. Hence, it is worth exploring the impact of higher-order biases and the non-linear effects, which plague these scales, on the clustering-redshifts method. We point out, however, that a conservative stand that includes only linear scales ($k_{\rm max} \sim 0.15\,\kunit$) is not strictly necessary, and in fact traditional applications of the clustering-redshifts approach using 2-point functions have made use of non-linear scales, e.g. down to separations of $\sim0.5\,{\rm Mpc}$ \citep{2019ApJ...877..150C}, showing robust constraints on the tomographically reconstructed quantities. Therefore, a target of $k_{\rm max} \sim 0.3\,\kunit$ is reasonable for clustering-redshifts calibration.

Based on these findings, we therefore recommend the use of the 3-point function to calibrate redshift distributions for future photometric datasets. A combination of interferometer and single-dish observations would likely be necessary in order to cover the whole redshift range out to $z\sim3$.

These conclusions are subject to a number of caveats. These are summarised as follows:
\begin{enumerate}
  \item We have neglected all non-Gaussian contributions to the bispectrum covariance. However, the off-diagonal correlations induced by the non-Gaussian terms is known to degrade the bispectrum constraints on primordial non-Gaussianity \citep{biagetti2021}. 
  \item We have only considered contributions from 2- and 3-point functions involving a single galaxy overdensity map, ignoring the information contained in the galaxy auto-spectrum and any other bispectrum combinations.
  \item We have assumed that the galaxy and HI biases are known exactly, including their evolution in redshift. While this may not be true in practice, this problem is inherent in the clustering-redshifts approach, and not specific to 21-cm data.
  \item We extrapolated the empirical second-order bias relation obtained for haloes, Eq. \ref{eq:b2}, to galaxies and neutral hydrogen, even though we should not expect this relation to hold for luminous tracers. Furthermore, we have not included any tidal, non-local, or higher-order bias terms\footnote{The impact of such biases -- and their marginalisation, as mentioned in item (iii) -- is an interesting issue that should be explored. Indeed, uncertainties on the details of the non-linear and scale-dependent galaxy biases should also affect the usual clustering-redshifts approach, and a more thorough investigation of their impact, or of the scale cuts needed to avoid them, is left for future work.}.
  \item We have made specific choices when modelling the impact of foreground contamination and photometric redshift uncertainties. Although these have been guided by achieved or forecast performance, the validity of the models used will depend strongly on e.g. complex instrumental effects, or the availability of the photo-$z$ training method used.
  \item Our forecasts have made use of the plane-parallel approximation.
\end{enumerate}
In summary, these conclusions must be treated as a proof-of-concept study, and an order-of-magnitude estimation of the information content of the 21cm-galaxy bispectrum.

It is worth comparing these results with those found in \citet{alonso2017} (A17), who studied the feasibility of clustering redshifts with 21 cm using the standard two-point function approach. In A17, foreground contamination was modelled in terms of a frequency decorrelation parameter $\xi$. It was found that, even for values of $\xi\sim0.1$, corresponding to a physical scale similar to the foreground cutoff used here $\kFG\sim0.01\,\kunit$, it would be possible to calibrate redshift distributions to the LSST requirement. This treatment of foregrounds is similar to the SM model of Eq. \ref{eq:FG_damping}, which effectively assumes that, although foregrounds dominate on scales below $\kFG$, the remaining cosmological signal is still accessible. The results found here agree with A17 when neglecting any foreground contamination ($\kFG=0$), and also qualitatively when using the SM foreground suppression model, similar in spirit to the decorrelation scale model used in A17. The main difference with respect to A17 is therefore the more conservative treatment of foregrounds used here, discarding all information below $\kFG$, and the inclusion of the horizon wedge.

Given the potential applicability of this method, the next step should be the development of robust estimators able to reconstruct the galaxy redshift distribution, as has been done for the standard clustering-redshifts approach \citep{mcquinn2013,1609.09085}, and their calibration against simulations with realistic levels of foreground contamination and photo-$z$ uncertainties. This should also include a thorough exploration of the different systematics affecting 21-cm observations, which could affect significantly the range of modes lost to foregrounds. We leave this for future work.

Finally, given the promising results presented here, it is worth thinking of other potential applications of higher-order correlations between 21-cm intensity mapping and photometric redshift surveys, or other projected probes of the large-scale structure (e.g. radio continuum surveys or CMB lensing maps). From a cosmological perspective, the motivation behind both of these probes is the same: overcoming the difficulty of measuring high-quality spectra for billions of galaxy in order to obtain high-sensitivity maps of the density inhomogeneities over large volumes. It would therefore be worth exploring whether these higher-order correlators are well-suited for the study of large-scale effects, such as the impact of primordial non-Gaussianity through its effects on the squeezed-limit bispectrum \citep{deputter2018}, or reconstructing super-sample density modes in general \citep{1804.02753,li2020}.

\section*{Acknowledgements}
We would like to thank Raul Abramo, Phil Bull, Steven Cunnington, and An\v{z}e Slosar for useful comments and discussions. We also thank the anonymous referee for comments that improved the final version of this work. CG is supported by the São Paulo Research Foundation (FAPESP), grant \href{https://bv.fapesp.br/pt/bolsas/180562/tecnicas-estatisticas-para-levantamentos-futuros-extraindo-fisica-primordial-da-estrutura-em-larga/?q=2018/10396-2}{2018/10396-2}. DA acknowledges support from the Beecroft Trust, and the Science and Technology Facilities Council through an Ernest Rutherford Fellowship, grant reference ST/P004474. KM acknowledges support from the National Research Foundation of South Africa. This work made extensive use of the public code \href{https://github.com/lesgourg/class_public}{\texttt{CLASS}} \citep{blas2011}, and the following python packages and libraries: \href{https://numpy.org/}{\texttt{numpy}} \citep{harris2020}, \href{https://scipy.org/}{\texttt{scipy}} \citep{2020SciPy}, \href{https://matplotlib.org/}{\texttt{matplotlib}} \citep{hunter2007}, and \href{https://github.com/LSSTDESC/CCL}{\texttt{pyccl}} \citep{chisari2019}. We thank Eline Maaike de Weerd for making her \href{https://github.com/ElineMaaikedeWeerd/bispectrum_snr}{SNR code} public, on which we based ourselves to compute the summation of Eq. \ref{eq:SNR}.

\section*{Data Availability}

The data and software used in this work are publicly available at \href{https://github.com/cmguandalin/ClusteringZ-21cm-Bispectrum}{https://github.com/cmguandalin/ClusteringZ-21cm-Bispectrum}.



\bibliographystyle{mnras}
\begin{multicols}{2}
\bibliography{main}
\end{multicols}



\appendix

\section{Clustering redshifts from 2- and 3-point functions}\label{app:math}
  \subsection{Preliminaries}\label{app:math.prelim}
    We will carry out most of our calculations in Fourier space. We use the following conventions for Fourier transforms in $N$ dimensions: 
    \begin{equation}
      f({\bs k})=\int {\rm d}^N{\bs r}\, f({\bs r})\,{\rm e}^{-i{\bs k}\cdot{\bs r}},
      \hspace{12pt}
      f({\bs r})=\int\frac{{\rm d}^N{\bs k}}{(2\pi)^N}\,f({\bs k})\,{\rm e}^{i{\bs k}\cdot{\bs r}}.
    \end{equation}

    Let us parametrize the selection function as a linear combination of basis functions $W_a(\apar{r})$:
    \begin{equation}
      \phi(\apar{r})=\sum_a \phi_a\,W_a(\apar{r}),\label{eq:decomposition}
    \end{equation}
    in which case the projected overdensity $\Delta_g$ can be written as a sum over ``slices'' $\delta_g^a$:
    \begin{equation}
      \Delta_g(\aperp{r})=\sum_a \phi_a\,\delta_g^a(\aperp{r}),\hspace{12pt}
      \delta_g^a(\aperp{r})\equiv\int {\rm d}\apar{r}W_a(\apar{r})\,\delta_g(\apar{r},\aperp{r}).
    \end{equation}
    Our aim is to quantify the accuracy with which the coefficients $\phi_a$ can be determined through cross-correlations.

    There are two obvious classes of basis functions that can be considered. The first is the \textbf{real-space top-hat}. Let $L$ be the range of radial comoving distance covered by the selection function. In this case, $W_a$ is a set of $N$ top-hat functions with width $\Delta x\equiv L/N$, centred at $\apar{r}^a\equiv(a+1/2)\Delta x$, with $a\in[0,N)$ and unit area. One can also consider the \textbf{Fourier decomposition}. In this second case, the basis functions are characterised by a radial wavenumber $\apar{k}$ and are given by:
      \begin{equation}
        W_{\apar{k}}(\apar{r})=\exp(-i\apar{k}\apar{r}).
      \end{equation}
      The Fourier basis is more manageable as the slices are simply the 3D Fourier coefficients of the galaxy overdensity $\delta_g^{\apar{k}}(\aperp{k})\equiv\delta_g(\apar{k},\aperp{k})$, and the selection function coefficients $\phi_{\apar{k}}$, in this basis, are related to their radial Fourier transform via
      \begin{equation}
        \phi_{\apar{k}}=\frac{\Delta \apar{k}}{2\pi}\phi(\apar{k})=\frac{1}{L}\phi(\apar{k}),
      \end{equation}
      where $\Delta\apar{k}\equiv2\pi/L$ is the natural Fourier spacing. Note that we assume throughout that the redshift support of each redshift bin is narrow enough to discard any evolution effects within it.

    \begingroup
    \allowdisplaybreaks
    We will consider 2- and 3-point correlators between the projected fields ($\Delta_g$, which we will label with a slice index $T$, and the slices $\delta_h^a$):
    \begin{equation}
      \langle \delta_x^a(\aperp{k})\delta_y^b(\aperp{q})\rangle\equiv(2\pi)^2\delta^D(\aperp{k}+\aperp{q}){\cal P}^{ab}_{xy}(k_\perp),\hspace{12pt}
      \langle \delta_x^a(\aperp{k})\delta_y^b(\aperp{q})\delta_z^c(\aperp{p})\rangle\equiv(2\pi)^2\delta^D(\aperp{k}+\aperp{q}+\aperp{p}){\cal B}^{abc}_{xyz}(k_\perp,q_\perp,p_\perp),\label{eq:projected_correlators}
    \end{equation}
    where ${\cal P}$ and ${\cal B}$ are the 2D power spectra and bispectra, respectively, and $x,y,z = \lbrace g, h\rbrace$ will label the type of tracer under consideration. The 3D analogues of these quantities, $P$ and $B$, are
    \begin{equation}
      \langle \delta_x({\bs k})\delta_y({\bs q})\rangle\equiv(2\pi)^3\delta^D({\bs k}+{\bs q})P_{xy}(k),\hspace{12pt}
      \langle \delta_x({\bs k})\delta_y({\bs q})\delta_z({\bs p})\rangle\equiv(2\pi)^3\delta^D({\bs k}+{\bs q}+{\bs p})B_{xyz}(k,q, p),
    \end{equation}where $k = |\bs{k}|$ is the modulus of the 3D Fourier vector. Using the Fourier radial decomposition, we can relate the different possible projected correlators, Eq. \ref{eq:projected_correlators}, with their 3D counterparts:
    \begin{align}
      &{\cal P}^{\apar{k}\apar{q}}_{xy}(k_\perp)=L\,\,\delta^K_{\apar{k}+\apar{q}}\,P_{xy}(k),\hspace{10pt}
      {\cal P}^{T\apar{q}}_{gx}(k_\perp)=\sum_{\apar{k}}L\,\,\delta^K_{\apar{k}+\apar{q}}\phi_{\apar{k}}P_{gx}(k),\hspace{10pt}
      {\cal P}^{TT}_{gg}(k_\perp)=\sum_{\apar{k}\apar{q}}L\,\,\delta^K_{\apar{k}+\apar{q}}\phi_{\apar{k}}\phi_{\apar{q}}P_{gg}(k),\label{eq:2d3d_power}\\
      &\text{ and }\nonumber \\
      &{\cal B}^{\apar{k}\apar{q}\apar{p}}_{xyz}(k_\perp,q_\perp,p_\perp)=L\,\,\delta^K_{\apar{k}+\apar{q}+\apar{p}}B_{xyz}(k,q,p),\hspace{10pt}
      {\cal B}^{T\apar{q}\apar{p}}_{gxy}(k_\perp,q_\perp,p_\perp)=\sum_{\apar{k}}L\,\,\delta^K_{\apar{k}+\apar{q}+\apar{p}}\phi_{\apar{k}}B_{gxy}(k,q,p),\label{eq:2d3d_bisp}\\
      &{\cal B}^{TT\apar{p}}_{ggx}(k_\perp,q_\perp,p_\perp)=\sum_{\apar{k}\apar{q}}L\,\,\delta^K_{\apar{k}+\apar{q}+\apar{p}}\phi_{\apar{k}}\phi_{\apar{q}}B_{ggx}(k,q,p),\hspace{10pt}
      {\cal B}^{TTT}_{ggg}(k_\perp,q_\perp,p_\perp)=\sum_{\apar{k}\apar{q}\apar{p}}L\,\,\delta^K_{\apar{k}+\apar{q}+\apar{p}}\phi_{\apar{k}}\phi_{\apar{q}}\phi_{\apar{p}}B_{ggg}(k,q,p).\nonumber
    \end{align}
    In these equations $\delta^D$ is the Dirac ``delta'' function, and $\delta^K_{k}$ is the Kronecker ``delta'' symbol, equal to 1 if $k=0$ and 0 otherwise.
    \endgroup
    
  \subsection{Fisher matrix}\label{ssec:fisher}
    The Fisher matrix for a data vector made up of a set of power spectra ${\cal P}^{XY}(k_\perp)$ is given by:
    \begin{equation}
      F^P_{\alpha\beta}=\sum_{{\bs X}{\bs X}'}\frac{A}{4\pi}\int_0^\infty {\rm d}k_\perp\,k_\perp\,\partial_\alpha {\cal P}^{XY}(k_\perp)\,\partial_\beta {\cal P}^{X'Y'}(k_\perp)\,{\cal I}^{XX'}(k_\perp)\,{\cal I}^{YY'}(k_\perp),\label{eq:power_spec_fisher}
    \end{equation}
    where ${\cal I}^{XY}$ is the $XY$ element of the inverse matrix of ${\cal P}^{XY}$, and ${\bs X} = \lbrace X, Y\rbrace$ runs over $T$ and all possible HI slices. Likewise for bispectra \citep{2006PhRvD..74b3522S,yadav2007,chen2021}:
    \begin{equation}
      F^B_{\alpha\beta}=\sum_{{\bs X}{\bs X}'}\frac{A}{4\pi}\frac{1}{6}\int_0^\infty {\rm d}k_\perp\,{\rm d}q_\perp\,{\rm d}p_\perp\,\frac{k_\perp q_\perp p_\perp}{\pi^2A_T}\,\partial_\alpha {\cal B}^{XYZ}(k_\perp,p_\perp,q_\perp)\,\partial_\beta {\cal B}^{X'Y'Z'}(k_\perp,p_\perp,q_\perp)\,{\cal I}^{XX'}(k_\perp)\,{\cal I}^{YY'}(q_\perp)\,{\cal I}^{ZZ'}(p_\perp),
    \end{equation}
    where $A_T=\frac{1}{2}\sqrt{2k_\perp^2q_\perp^2+2q_\perp^2p_\perp^2+2p_\perp^2k_\perp^2-k_\perp^4-q_\perp^4-p_\perp^4}$ is the area of a triangle with sides $(k_\perp,p_\perp,q_\perp)$, $A$ is the survey area, and ${\bs X} = \lbrace X, Y, Z\rbrace$.
 
    In our case, the only parameters we will consider are the selection function coefficients $\phi_{\apar{k}}$. Let us denote $\partial_{\apar{k}}\equiv\partial/\partial\phi_{\apar{k}}$; then, following Eqs. \ref{eq:2d3d_power} and \ref{eq:2d3d_bisp}, the only non-vanishing derivatives of the different power spectra and bispectra are:
    \begingroup
    \allowdisplaybreaks
    \begin{align*}
      \partial_{\apar{p}}{\cal P}^{T\apar{q}}_{gh}(k_\perp)&=L\,\,\delta^K_{\apar{p}+\apar{q}}P_{gh}(\apar{p},k_\perp),\\
      \partial_{\apar{p}}{\cal P}^{TT}_{gg}(k_\perp)&=2L\,\,\phi_{-\apar{p}}P_{gg}(\apar{p},k_\perp),\\
      \partial_{\apar{l}}{\cal B}^{T\apar{q}\apar{p}}_{ghh}(k_\perp,p_\perp,q_\perp)&=L\,\,\delta^K_{\apar{l}+\apar{q}+\apar{p}}B_{ghh}(\apar{l},\apar{q},\apar{p};\bs{\triangle}_{\perp}),\\
      \partial_{\apar{l}}{\cal B}^{TT\apar{p}}_{ggh}(k_\perp,p_\perp,q_\perp)&=L\,\,\phi_{-\apar{l}-\apar{p}}\left[B_{ggh}(\apar{l},-\apar{l}-\apar{p},\apar{p};\bs{\triangle}_{\perp})+B_{ggh}(-\apar{l}-\apar{p},\apar{l},\apar{p};\bs{\triangle}_{\perp})\right],\\
      \partial_{\apar{l}}{\cal B}^{TTT}_{ggh}(k_\perp,p_\perp,q_\perp)&=\sum_{\apar{p}}L\,\,\phi_{\apar{p}}\phi_{-\apar{l}-\apar{p}}\left[
      B_{ggg}(\apar{l},-\apar{l}-\apar{p},\apar{p};\bs{\triangle}_{\perp})+
      B_{ggg}(\apar{p},\apar{l},-\apar{l}-\apar{p};\bs{\triangle}_{\perp})+
      B_{ggg}(-\apar{l}-\apar{p},\apar{p},\apar{l};\bs{\triangle}_{\perp})\right],
  \end{align*} since $\partial_{\apar{p}}{\cal P}^{\apar{k}\apar{q}}_{hh}(k_\perp)=0$ and $\partial_{\apar{l}}{\cal B}^{\apar{k}\apar{p}\apar{q}}_{hhh}(k_\perp,p_\perp,q_\perp)=0$ as the autocorrelations of HI maps do not depend on $\phi_{\apar{k}}$. Above, we defined the short-hand notation $\bs{\triangle}_\perp\equiv(k_\perp,p_\perp,q_\perp)$. Finally, the inverse power-spectrum elements are given by:
  \begin{align}
    &{\cal I}^{TT}(k_\perp)=\left\{\sum_{\apar{k}}L\,|\phi_{\apar{k}}|^2P_{gg}(k)\left[1-\frac{P_{gh}^2(k)}{P_{gg}(k)P_{hh}(k)}\right]\right\}^{-1},\label{eq:ITT}\\
    &{\cal I}^{T\apar{k}}(k_\perp)=-{\cal I}^{TT}(k_\perp)\,\phi_{\apar{k}}\frac{P_{gh}(k)}{P_{hh}(k)},\label{eq:ITk}\\
    &{\cal I}^{\apar{k}\apar{q}}(k_\perp)=\frac{\delta^K_{\apar{k}+\apar{q}}}{L\,P_{hh}(k)}+{\cal I}^{TT}(k_\perp)\,\phi_{\apar{k}}\phi_{\apar{q}}\frac{P_{gh}(\apar{k},k_\perp)}{P_{hh}(\apar{k},k_\perp)}\frac{P_{gh}(\apar{q},k_\perp)}{P_{hh}(\apar{q},k_\perp)}\simeq\frac{\delta^K_{\apar{k}+\apar{q}}}{L\,P_{hh}(k)}.\label{eq:Ikq}
  \end{align}
  \endgroup

  \subsubsection{Power spectrum Fisher matrix}\label{sssec:fisher_p}
    The power spectrum Fisher matrix, Eq. \ref{eq:FisherP}, can be decomposed into the following four non-zero contributions:
    \begin{equation}
      F^{P}_{\apar{k}\apar{q}}=F^{2T2T}_{\apar{k}\apar{q}}+4F^{2T1T}_{(\apar{k}\apar{q})}+2F^{1T1T+}_{\apar{k}\apar{q}}+2F^{1T1T\times}_{\apar{k}\apar{q}},
    \end{equation}
    where $4F^{2T1T}_{(\apar{k}\apar{q})} = 2F^{2T1T}_{\apar{k}\apar{q}} +2F^{2T1T}_{\apar{q}\apar{k}}$ and
    \begingroup
    \allowdisplaybreaks
    \begin{align}\nonumber
      F^{2T2T}_{\apar{k}\apar{q}}
      &=\frac{A}{4\pi}\int {\rm d}k_\perp\,k_\perp\,\partial_{\apar{k}}{\cal P}^{TT}_{gg}(k_\perp)\,\partial_{\apar{q}}{\cal P}^{TT}_{gg}(k_\perp)\left[{\cal I}^{TT}(k_\perp)\right]^2\\
      &=\frac{A}{\pi}L^2\phi_{-\apar{k}}\phi_{-\apar{q}}\int {\rm d}k_\perp\,k_\perp\,{\cal R}_{gg}(\apar{k},k_\perp)\,{\cal R}_{gg}(\apar{q},k_\perp)\\\nonumber
      F^{2T1T}_{\apar{k}\apar{q}}
      &=\frac{A}{4\pi}\sum_{\apar{p}}\int {\rm d}k_\perp\,k_\perp\,\partial_{\apar{k}}{\cal P}^{TT}_{gg}(k_\perp)\,\partial_{\apar{q}}{\cal P}^{T\apar{p}}_{gh}(k_\perp)\,{\cal I}^{TT}(k_\perp)\,{\cal I}^{T\apar{p}}(k_\perp)\\
      &=-\frac{A}{4\pi}2L^2\phi_{-\apar{k}}\phi_{-\apar{q}}\int {\rm d}k_\perp\,k_\perp\,{\cal R}_{gg}(\apar{k},k_\perp)\,{\cal R}^2_{gh}(\apar{q},k_\perp)\\\nonumber
      F^{1T1T+}_{\apar{k}\apar{q}}
      &=\frac{A}{4\pi}\sum_{\apar{p}\apar{l}}\int {\rm d}k_\perp\,k_\perp\,\partial_{\apar{k}}{\cal P}^{T\apar{p}}_{gh}(k_\perp)\,\partial_{\apar{q}}{\cal P}^{T\apar{l}}_{gh}(k_\perp)\,{\cal I}^{TT}(k_\perp)\,{\cal I}^{\apar{p}\apar{l}}(k_\perp)\\
      &=\frac{AL}{4\pi}\delta^K_{\apar{k}+\apar{q}}\int {\rm d}k_\perp\,k_\perp{\cal R}_{gh}^2(\apar{k},k_\perp)+
      \frac{A}{4\pi}L^2\phi_{-\apar{k}}\phi_{-\apar{q}}\int {\rm d}k_\perp\,k_\perp{\cal R}_{gh}^2(\apar{k},k_\perp)\,{\cal R}^2_{gh}(\apar{q},k_\perp),\\\nonumber
      F^{1T1T\times}_{\apar{k}\apar{q}}
      &=\frac{A}{4\pi}\sum_{\apar{p}\apar{l}}\int {\rm d}k_\perp\,k_\perp\,\partial_{\apar{k}}{\cal P}^{T\apar{p}}_{gh}(k_\perp)\,\partial_{\apar{q}}{\cal P}^{T\apar{l}}_{gh}(k_\perp)\,{\cal I}^{T\apar{p}}(k_\perp)\,{\cal I}^{T\apar{l}}(k_\perp)\\
      &=\frac{A}{4\pi}L^2\phi_{-\apar{k}}\phi_{-\apar{q}}\int {\rm d}k_\perp\,k_\perp\,{\cal R}_{gh}^2(\apar{k},k_\perp)\,{\cal R}_{gh}^2(\apar{q},k_\perp)
    \end{align}
    where
    \begin{equation}
      {\cal R}_{gg}(\apar{k},k_\perp)\equiv P_{gg}(\apar{k},k_\perp)\,{\cal I}^{TT}(k_\perp),\hspace{12pt}{\cal R}^2_{gh}(\apar{k},k_\perp)\equiv\frac{P^2_{gh}(\apar{k},k_\perp)}{P_{hh}(\apar{k},k_\perp)}\,{\cal I}^{TT}(k_\perp).
    \end{equation}
    Putting everything together:
    \begin{align}\nonumber
      F^{P}_{\apar{k}\apar{q}}
      &=\frac{AL^2}{2\pi}\left[\frac{\delta^K_{\apar{k}+\apar{q}}}{L}\int {\rm d}k_\perp k_\perp {\cal R}^2_{gh}(\apar{k},k_\perp)+2\phi_{-\apar{k}}\phi_{-\apar{q}}\int {\rm d}k_\perp k_\perp \left({\cal R}_{gg}(\apar{k},k_\perp)+{\cal R}^2_{gh}(\apar{k},k_\perp)\right)\left({\cal R}_{gg}(\apar{q},k_\perp)+{\cal R}^2_{gh}(\apar{q},k_\perp)\right)\right]\\\label{eq:fisher_P_phi}
      &\simeq\delta^K_{\apar{k}+\apar{q}}\frac{AL}{2\pi}\int {\rm d}k_\perp k_\perp{\cal I}^{TT}(k_\perp)\frac{P_{gh}^2(\apar{k},k_\perp)}{P_{hh}(\apar{k},k_\perp)},
    \end{align}
    where the second equality holds to lowest order in $P_{gh}/P_{hh}$ and neglecting any information coming from ${\cal P}^{TT}_{gg}(k_\perp)$. The latter approximation also allows us to isolate the constraints coming only from the HI cross-correlation, avoiding self-calibration constraints that would be degenerate with e.g. uncertainties on the galaxy bias or the impact of systematic uncertainties in the galaxy autocorrelation.
    \endgroup

  \subsubsection{The bispectrum Fisher matrix}
    The bispectrum Fisher matrix can be decomposed into the following non-zero contributions:
    \begin{equation}
      F^B_{\apar{k}\apar{q}}=F^{3T3T}_{\apar{k}\apar{q}}+6F^{3T2T}_{(\apar{k},\apar{q})}+6F^{3T1T}_{(\apar{k},\apar{q})}+3F^{2T2T+}_{\apar{k}\apar{q}}+6F^{2T2T\times}_{\apar{k}\apar{q}}+12F^{2T1T+}_{(\apar{k},\apar{q})}+6F^{2T1T\times}_{(\apar{k},\apar{q})}+3F^{1T1T+}_{\apar{k}\apar{q}}+6F^{1T1T\times}_{\apar{k}\apar{q}},
    \end{equation}
    where, e.g. $6F^{3T2T}_{(\apar{k},\apar{q})} \equiv 3F^{3T2T}_{\apar{k},\apar{q}} + 3F^{3T2T}_{\apar{q},\apar{k}}$, and:
    \begingroup
    \allowdisplaybreaks
    \begin{align}
      F^{3T3T}_{\apar{k}\apar{q}}
      &=\int{\cal D}(k_\perp,q_\perp,p_\perp)\,
      \partial_{\apar{k}}B^{TTT}_{ggg}(\bs{\triangle}_\perp)\,\partial_{\apar{q}}B^{TTT}_{ggg}(\bs{\triangle}_\perp)\,{\cal I}^{TT}(k_\perp)\,{\cal I}^{TT}(q_\perp)\,{\cal I}^{TT}(p_\perp),\\
      F^{3T2T}_{\apar{k}\apar{q}}
      &=\sum_{\apar{p}}\int{\cal D}(k_\perp,q_\perp,p_\perp)\,
      \partial_{\apar{k}}B^{TTT}_{ggg}(\bs{\triangle}_\perp)\,\partial_{\apar{q}}B^{TT\apar{p}}_{ggh}(\bs{\triangle}_\perp)\,{\cal I}^{TT}(k_\perp)\,{\cal I}^{TT}(q_\perp)\,{\cal I}^{T\apar{p}}(p_\perp),\\
      F^{3T1T}_{\apar{k}\apar{q}}&=\sum_{\apar{p}\apar{r}}\int{\cal D}(k_\perp,q_\perp,p_\perp)\,\partial_{\apar{k}}B^{TTT}_{ggg}(\bs{\triangle}_\perp)\,\partial_{\apar{q}}B^{T\apar{p}\apar{r}}_{ghh}(\bs{\triangle}_\perp)\,{\cal I}^{TT}(k_\perp)\,{\cal I}^{T\apar{p}}(q_\perp)\,{\cal I}^{T\apar{r}}(p_\perp),\\
      F^{2T2T+}_{\apar{k}\apar{q}}&=\sum_{\apar{p}\apar{r}}\int{\cal D}(k_\perp,q_\perp,p_\perp)\,\partial_{\apar{k}}B^{TT\apar{p}}_{ggh}(\bs{\triangle}_\perp)\,\partial_{\apar{q}}B^{TT\apar{r}}_{ggh}(\bs{\triangle}_\perp)\,{\cal I}^{TT}(k_\perp)\,{\cal I}^{TT}(q_\perp)\,{\cal I}^{\apar{p}\apar{r}}(p_\perp),\\
      F^{2T2T\times}_{\apar{k}\apar{q}}&=\sum_{\apar{p}\apar{r}}\int{\cal D}(k_\perp,q_\perp,p_\perp)\,\partial_{\apar{k}}B^{TT\apar{p}}_{ggh}(\bs{\triangle}_\perp)\,\partial_{\apar{q}}B^{T\apar{r}T}_{ghg}(\bs{\triangle}_\perp)\,{\cal I}^{TT}(k_\perp)\,{\cal I}^{T\apar{r}}(q_\perp)\,{\cal I}^{T\apar{p}}(p_\perp),\\
      F^{2T1T+}_{\apar{k}\apar{q}}&=\sum_{\apar{p}\apar{r}\apar{s}}\int{\cal D}(k_\perp,q_\perp,p_\perp)\,\partial_{\apar{k}}B^{TT\apar{p}}_{ggh}(\bs{\triangle}_\perp)\,\partial_{\apar{q}}B^{T\apar{r}\apar{s}}_{ghh}(\bs{\triangle}_\perp)\,{\cal I}^{TT}(k_\perp)\,{\cal I}^{T\apar{r}}(q_\perp)\,{\cal I}^{\apar{p}\apar{s}}(p_\perp),\\
      F^{2T1T\times}_{\apar{k}\apar{q}}&=\sum_{\apar{p}\apar{r}\apar{s}}\int{\cal D}(k_\perp,q_\perp,p_\perp)\,\partial_{\apar{k}}B^{TT\apar{p}}_{ggh}(\bs{\triangle}_\perp)\,\partial_{\apar{q}}B^{\apar{r}\apar{s}T}_{hhg}(\bs{\triangle}_\perp)\,{\cal I}^{T\apar{r}}(k_\perp)\,{\cal I}^{T\apar{s}}(q_\perp)\,{\cal I}^{T\apar{p}}(p_\perp),\\
      F^{1T1T+}_{\apar{k}\apar{q}}&=\sum_{\apar{p}\apar{r}\apar{s}\apar{t}}\int{\cal D}(k_\perp,q_\perp,p_\perp)\,\partial_{\apar{k}}B^{T\apar{p}\apar{r}}_{ggg}(\bs{\triangle}_\perp)\,\partial_{\apar{q}}B^{T\apar{s}\apar{t}}_{ggg}(\bs{\triangle}_\perp)\,{\cal I}^{TT}(k_\perp)\,{\cal I}^{\apar{p}\apar{s}}(q_\perp)\,{\cal I}^{\apar{r}\apar{t}}(p_\perp),\\
      F^{1T1T\times}_{\apar{k}\apar{q}}
      &=\sum_{\apar{p}\apar{r}\apar{s}\apar{t}}\int{\cal D}(k_\perp,q_\perp,p_\perp)\,\partial_{\apar{k}}B^{T\apar{p}\apar{r}}_{ggg}(\bs{\triangle}_\perp)\,\partial_{\apar{q}}B^{\apar{s}T\apar{t}}_{ggg}(\bs{\triangle}_\perp)\,{\cal I}^{T\apar{s}}(k_\perp)\,{\cal I}^{T\apar{p}}(q_\perp)\,{\cal I}^{\apar{r}\apar{t}}(p_\perp),
    \end{align}
    \endgroup
    with
    \begin{equation}
      {\cal D}(k_\perp,q_\perp,p_\perp)\equiv\frac{A}{4\pi}\frac{1}{6}\,{\rm d}k_\perp {\rm d}q_\perp {\rm d}p_\perp\, \frac{k_\perp q_\perp p_\perp}{\pi^2A_T}.
    \end{equation}

    As before, to lowest order in $P_{gh}/P_{hh}$, and ignoring any information from $B^{TTT}$ or $B^{TT\apar{k}}$ (i.e. keeping only data combinations that are linear in the unknown parameters $\phi_{\apar{k}}$), the only non-zero component is $F^{1T1T+}_{\apar{k}\apar{q}}$ which, in this approximation, becomes:
    \begin{equation}\label{eq:fisher_B_phi}
      F^{B}_{\apar{k}\apar{q}}\simeq\delta^K_{\apar{k}+\apar{q}}\frac{AL}{4\pi}\frac{1}{6}\int {\rm d}k_\perp {\rm d}q_\perp {\rm d}p_\perp\frac{k_\perp q_\perp p_\perp}{\pi^2A_T}\,{\cal I}^{TT}(k_\perp)\int\frac{{\rm d}\apar{p}}{2\pi}\frac{B^2_{ghh}(\apar{k},\apar{p},-\apar{k}-\apar{p},\bs{\triangle}_\perp)}{P_{hh}(\apar{p},q_\perp)P_{hh}(-\apar{k}-\apar{p},p_\perp)}.
    \end{equation}

  \subsubsection{Parametrizing redshift uncertainty}\label{app:gaussian}
    To condense the clustering redshift constraints into a small number of parameters for which the forecast uncertainties can be compared with existing requirements for future photometric redshifts, let us model the redshift distribution $\phi(\apar{r})$ as a simple Gaussian centred at $\apar{r}^*$ with standard deviation $\apar{\sigma}$, treating both as free parameters. The Fourier coefficients of this distribution are
    \begin{equation}
      \phi_{\apar{k}}=\frac{1}{L}{\rm e}^{-\frac{\apar{k}^2\apar{\sigma}^2}{2}}{\rm e}^{-i\apar{k}\apar{r}^*}.
    \end{equation}
    
    It is then possible to propagate the Fisher matrix of the redshift distribution amplitudes $\phi_{\apar{k}}$ (Eqs. \ref{eq:fisher_P_phi} and \ref{eq:fisher_B_phi} for the power spectrum and bispectrum respectively) into that of the two free parameters $\apar{r}^*$ and $\apar{\sigma}$:
    \begin{equation}
      F_{\theta_1\theta_2}=\sum_{\apar{k}\apar{q}}\partial_{\theta_1}\phi_{\apar{k}}\partial_{\theta_2}\phi_{\apar{q}}F_{\apar{k}\apar{q}}.
    \end{equation}
    Doing so, and after taking the continuum limit in all radial wavenumbers ($\apar{k}$ etc.) we obtain the following result:

    \paragraph*{For the power spectrum:}
      \begin{align}
        {\rm Var}^{-1}_P(\apar{r}^*)&=\frac{A}{2\pi} \int {\rm d}k_\perp k_\perp \, {\cal I}^{TT}(k_{\perp})\int\frac{\mathrm{d}\apar{k}}{2\pi}\apar{k}^2\,\mathrm{e}^{-\apar{k}^2\apar{\sigma}^2} \frac{P^2_{gh}(\apar{k},k_\perp)}{P_{hh}(\apar{k},k_\perp)} - {\cal C}_{\apar{r}^*}(\apar{k},\apar{q},k_\perp),
      \end{align}
      \begin{align}
        {\rm Var}^{-1}_P(\apar{\sigma})&=\frac{A}{2\pi} \int {\rm d}k_\perp k_\perp \, {\cal I}^{TT}(k_{\perp})\int\frac{{\rm d}\apar{k}}{2\pi}\apar{\sigma}^2\apar{k}^4\,{\rm e}^{-\apar{k}^2\apar{\sigma}^2} \frac{P^2_{gh}(\apar{k},k_\perp)}{P_{hh}(\apar{k},k_\perp)} + {\cal C}_{\apar{\sigma}}(\apar{k},\apar{q},k_\perp),
      \end{align}
      where
      \begin{equation}
          {\cal C}_{\theta}(\apar{k},\apar{q},k_\perp) = \frac{A}{\pi} \int {\rm d}k_\perp k_\perp \left[{\cal I}^{TT}(k_{\perp})\right]^2\Bigg\lbrace \int\frac{ \mathrm{d}\apar{k}}{2\pi}\,\frac{\mathrm{d}\apar{q}}{2\pi}\,\,{\cal K}_\theta(k_\parallel,\sigma_\parallel)\,{\cal K}_\theta(q_\parallel,\sigma_\parallel)\,\mathrm{e}^{-\apar{\sigma}^2(\apar{k}^2+\apar{q}^2)}\, {\cal R}(\apar{k},k_\perp)\, {\cal R}(\apar{q},k_\perp)\Bigg\rbrace,
      \end{equation}with
      \begin{equation}
        {\cal R}(\apar{k},k_\perp) \equiv P_{gg}(\apar{k},k_\perp)-\frac{P^2_{gh}(\apar{k},k_\perp)}{P_{hh}(\apar{k},k_\perp)},\hspace{5mm} {\cal K}_{\theta = r^*_\parallel} = \apar{k}, \hspace{5mm}{\cal K}_{\theta = \sigma^*_\parallel} = \sigma_\parallel\,\apar{k}^2.
      \end{equation}While these corrections overconstrain $\theta = \apar{\sigma}$ for the reasons already mentioned, they are negligible for $\theta = \apar{r}^*$. Therefore, we neglect them for the remaining of this work, and employ Eqs. \ref{eq:var_r_P} and \ref{eq:var_s_P} for the power-spectrum constraints shown in Section \ref{sec:results} and Appendix \ref{ap:extra}.
    \paragraph*{For the bispectrum:}
      \begin{align}
        &{\rm Var}^{-1}_B(\apar{r}^*)\simeq\frac{1}{6}\frac{A}{4\pi}\int {\rm d}k_\perp {\rm d}p_\perp {\rm d}q_\perp \frac{k_\perp p_\perp q_\perp}{\pi^2A_T}\,{\cal I}^{TT}(k_\perp)\int\frac{{\rm d}\apar{p}}{2\pi}\frac{{\rm d}\apar{k}}{2\pi}\apar{k}^2{\rm e}^{-\apar{k}^2\apar{\sigma}^2}\frac{B^2_{ghh}(\apar{k},\apar{p},-\apar{k}-\apar{p};k_\perp,p_\perp,q_\perp)}{P_{hh}(\apar{p},q_\perp)P_{hh}(-\apar{k}-\apar{p},p_\perp)},\label{eq:FisherB_r}\\
        &{\rm Var}^{-1}_B(\apar{\sigma}^*)\simeq\frac{1}{6}\frac{A}{4\pi}\int {\rm d}k_\perp {\rm d}p_\perp {\rm d}q_\perp \frac{k_\perp p_\perp q_\perp}{\pi^2A_T}\,{\cal I}^{TT}(k_\perp)\int\frac{{\rm d}\apar{p}}{2\pi}\frac{{\rm d}\apar{k}}{2\pi}\apar{\sigma}^2\apar{k}^4{\rm e}^{-\apar{k}^2\apar{\sigma}^2}\frac{B^2_{ghh}(\apar{k},\apar{p},-\apar{k}-\apar{p};k_\perp,p_\perp,q_\perp)}{P_{hh}(\apar{p},q_\perp)P_{hh}(-\apar{k}-\apar{p},p_\perp)},\label{eq:FisherB_s}
      \end{align}
    and ${\rm Cov}(\apar{r}^*,\apar{\sigma})=0$ in both cases, in agreement with Eqs. \ref{eq:var_r_P}, \ref{eq:var_s_P}, \ref{eq:var_r_B}, and \ref{eq:var_s_B}. Notice that, in Eqs. \ref{eq:FisherB_r} and \ref{eq:FisherB_s}, we are already neglecting the galaxy autocorrelations and high-order corrections as in Eq. \ref{eq:fisher_B_phi}.

\begin{figure}
    \centering
    \includegraphics[width=\columnwidth]{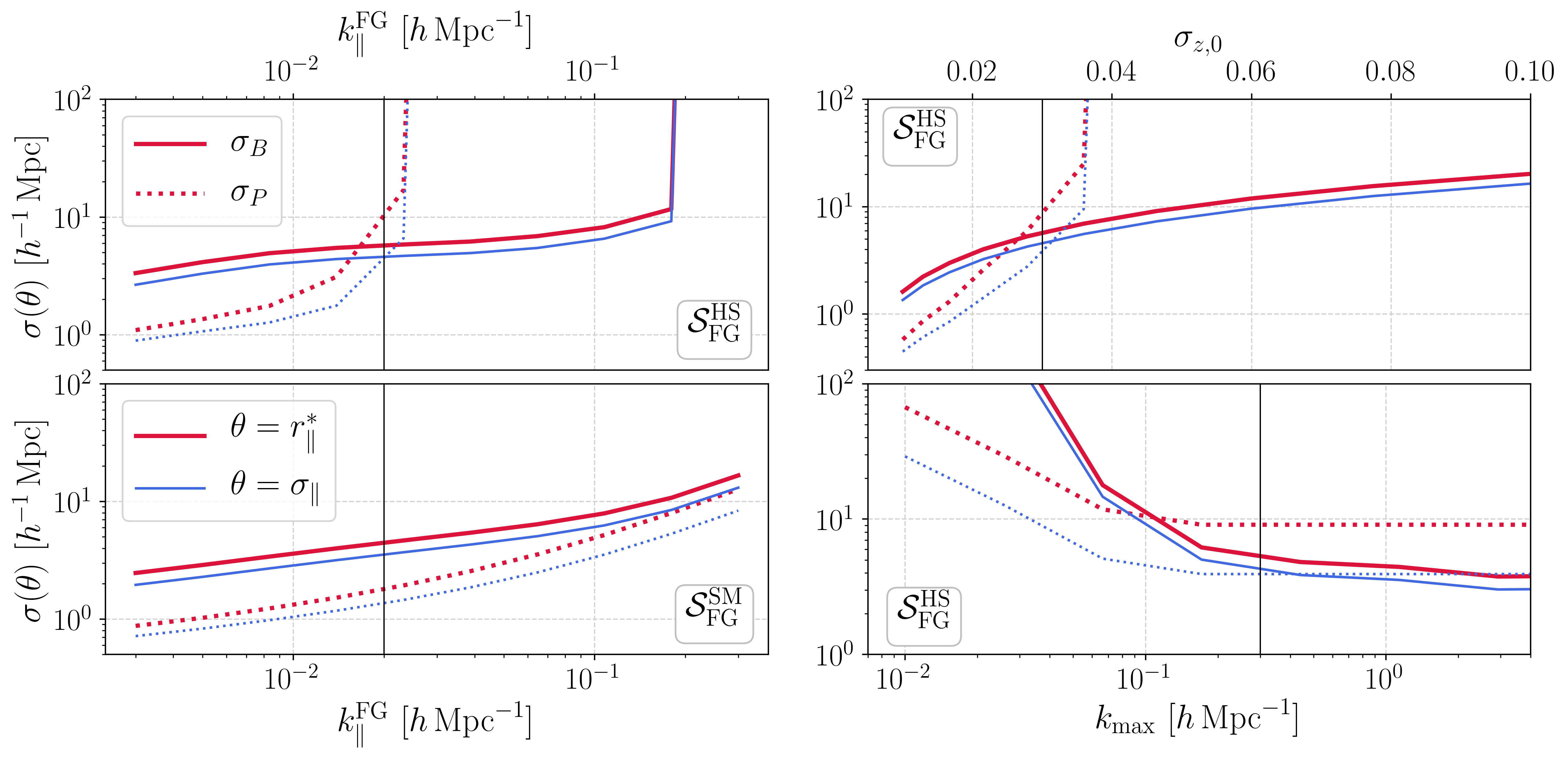}
    \caption{As in Fig. \ref{fig:HIRAX_idealised}, we show the power-spectrum (\textit{dotted}) and bispectrum (\textit{solid}) constraints obtained by cross-correlating an SKA-like (single-dish) survey and a LSST-like, on the two Gaussian photo-$z$ parameters, $\theta = \apar{r}^*$ (\textit{thick red}) and $\theta = \apar{\sigma}$ (\textit{thin blue}). The dependences on the foreground cutoff/damping scale $\apar{k}^{\rm FG}$ ({\it left, top and bottom}), photo-$z$ scatter $\sigma_{z,0}$ ({\it right, top}), and maximum wavenumber $k_{\rm max}$ ({\it right, bottom}) are shown assuming no second-order bias correction for the bispectrum. In the bottom-left panel, foregrounds are treated with the smooth exponential damping, ${\cal S}_{\rm FG}^{\rm SM}(\apar{k})$ in Eq. \ref{eq:FG_damping}. We take as the fiducial modelling for the other plots the Heaviside model, ${\cal S}_{\rm FG}^{\rm HS}(\apar{k})$, by fixing $\apar{k}^{\rm min} = \apar{k}^{\rm FG}$. Vertical lines mark the fiducial values used in our analyses.}
    \label{fig:SKA_idealised}
\end{figure}

\section{Additional SKA analysis}\label{ap:extra}

In this appendix we extend the results presented in Section \ref{ssec:simplified_results} for an SKA-like survey (Fig. \ref{fig:SKA_idealised}). We consider the same redshift bin, centred at $z = 0.8$, with fixed linear bias $b \approx 1.4$ and number density $\bar{n} \approx 0.004$ for the photometric galaxies, and explore how the cross-correlations depend on the width of the Gaussian kernel $\sigma_{z,0}$, foreground cutoff scale $k_{\rm FG}$ and maximum wavenumber of integration $k_{\rm max}$.

For the SKA survey, the beam attenuation (as modelled by ${\cal S}_b$) is a relevant feature, damping the high-$k_\perp$ modes as in Fig. \ref{fig:beam_FG}. As a consequence, we are unable to access a larger number of perpendicular modes deep inside the non-linear regime, where the bispectrum amplitude is larger, if compared to the interferometer mode. In this case, the bispectrum constraints are degraded and they cannot outperform the power-spectrum method for the smooth foreground removal, as opposed to Fig. \ref{fig:HIRAX_idealised}. For the conservative foreground cutoff model, the qualitative results remain: whilst there is a divergence around $\apar{k}^{\rm FG} \sim 0.02\, \kunit$ for the power spectrum, with the bispectrum we can push the clustering-redshifts method up to $\apar{k}^{\rm FG} \sim 0.2\, \kunit$. 

When it comes to the effects of the other parameters, the bispectrum is well behaved as a function of the Gaussian width $\sigma_{z,0}$, and we also observe a convergence with respect to the maximum scale of integration, at around $k_{\rm max} \sim 1\,\kunit$. 

\section{Signal-to-noise ratio}\label{ap:snr}
As mentioned in Section \ref{ssec:stageIV}, at higher redshifts the signal from the bispectrum becomes weaker as the density field becomes more Gaussian. It is possible to quantify this effect by looking at the bispectrum signal-to-noise ratio (SNR) as a function of redshift \citep{scoccimarro2004,maartens2020}:
\begin{equation}
    \left[\frac{S}{N}(z)\right]^2 = \sum_{k_1,k_2,k_3} \frac{B^2(z;k_1,k_2,k_3)}{\mathrm{Var}[\hat{B}(z;k_1,k_2,k_3)]}, \,\,\,\, \text{where}\,\, \sum_{k_1,k_2,k_3} = \sum_{k_1=k_{\rm min}}^{k_{\rm max}} \sum_{k_2=k_{\rm min}}^{k_1} \sum_{k_3=k_{\rm min}}^{k_2}. \label{eq:SNR}
\end{equation}

Under the Gaussian assumption for the bispectrum covariance \citep{chan2017}, 
\begin{align}
    \mathrm{Var}[\hat{B}(z;k_1,k_2,k_3)] = \frac{(2\pi)^3}{V}\frac{s_{123}P(k_1)P(k_2)P(k_3)}{8\pi^2 k_1k_2k_3(\Delta k)^3\beta(\triangle)},
\end{align}where $V$ is the survey volume, $s_{123} = 6, 2, 1$ for equilateral, isosceles and scalene triangles, respectively, and 
\begin{equation}
    \triangle = \frac{k_3^2-k_1^2-k_2^2}{2k_1k_2}, \,\,\,\, \beta(\triangle) = \begin{cases}0.5,& \triangle = \pm 1,\\ 1, & 0< \triangle <1\\ 0, & \text{otherwise} \end{cases}.
\end{equation}

Therefore:
\begin{align}
    \left[\frac{S}{N}(z)\right]^2 &= \frac{V}{\pi}\sum_{k_1,k_2,k_3} k_1k_2k_3(\Delta k)^3 \frac{\beta(\triangle)}{s_{123}} \frac{B^2(z;k_1,k_2,k_3)}{P(z;k_1)P(z;k_2)P(z;k_3)},\label{eq:SNRmatter}
\end{align}where we took $\Delta k$ to be the fundamental mode of the survey, i.e. $\Delta k \equiv k_f = 2\pi/V^{1/3}$.

\begin{figure}
    \centering
    \includegraphics[width=0.5\columnwidth]{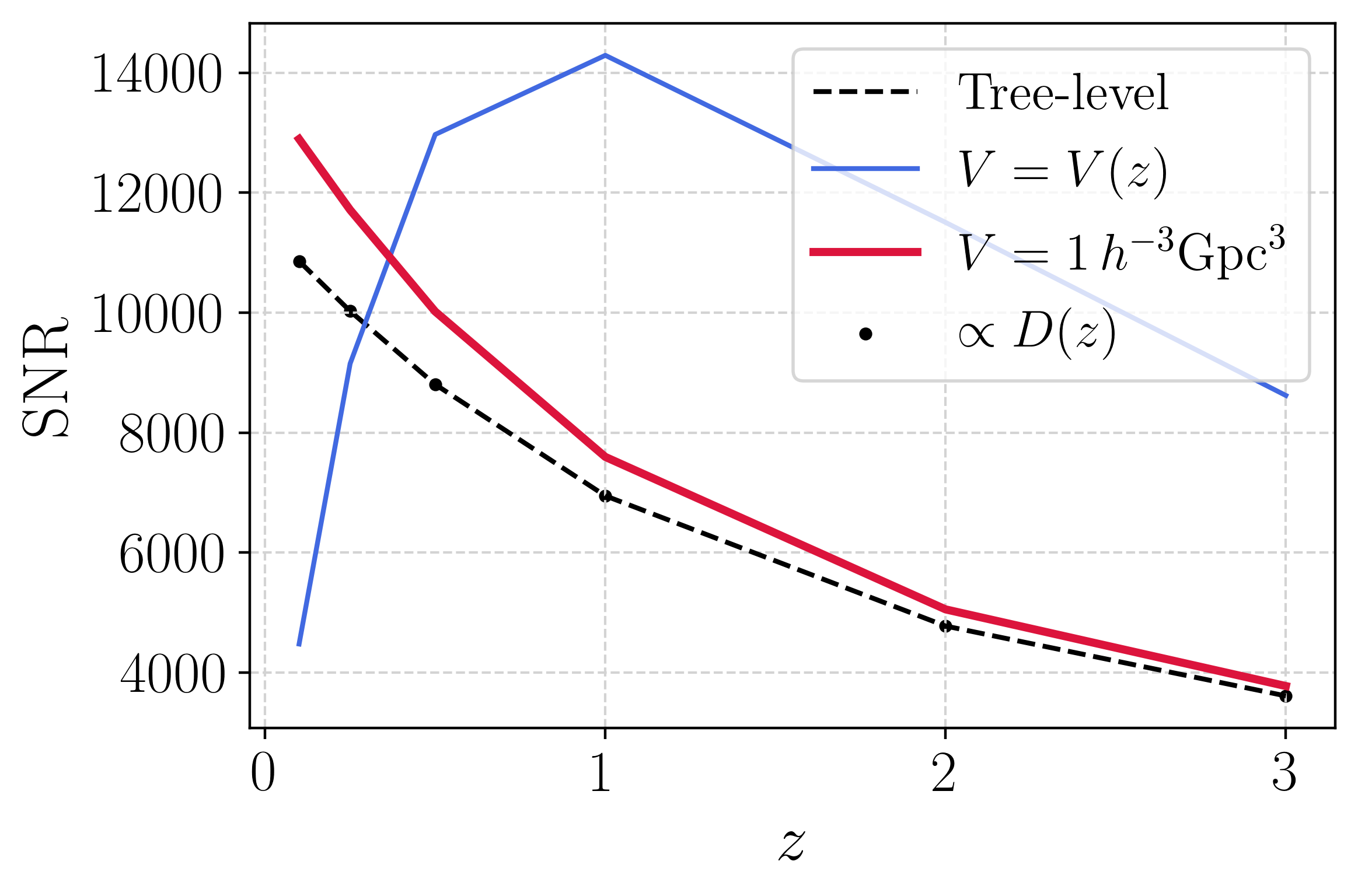}
    \caption{Signal-to-noise ratio (SNR) for the matter bispectrum, as given by Eq. \ref{eq:SNRmatter}. For the redshift-dependent case (\textit{thin blue}), we model the signal via Eq. \ref{eq:bispectrumFeff}. For this case, as the volume for each redshift bin increases ($z<1$), the SNR becomes larger; however, for $z\geq 1$, the decrease in the bispectrum amplitude is more relevant and causes the SNR to drop. For the fixed survey volume, however, there is a monotonic decrease in the SNR proportional to the growth factor $D(z)$ (\textit{dots}), as $B^2/P^3 \sim D^2(z)$ for the bispectrum induced by gravity at tree-level (\textit{black-dashed} curve), which scales as $P^2(k)$. For the signal given by Eq. \ref{eq:bispectrumFeff} (\textit{thick red}), there are deviations from a pure $D(z)$ scaling, since terms beyond the linear growth must be considered. We can also see that the tree-level limit underestimates the amplitude of the signal, especially at lower redshifts where non-linearities are stronger.}
    \label{fig:snr}
\end{figure}

The SNR of the matter bispectrum, Eq. \ref{eq:SNRmatter}, is shown in Fig. \ref{fig:snr} for the cases in which the survey volume is fixed to $V = 1\,h^{-3}\,\,{\rm Gpc}^3$, and when the volume is redshift dependent: $V(z) = \frac{4\pi f_\mathrm{sky}}{3}(r_+^3 - r_-^3)$, with $r_{\rm \pm} = \chi(\bar{z} \pm \frac{\Delta z}{2})$ (assuming bins of $\Delta z = 0.1$ centred at $\bar{z}$), and $4\pi f_{\rm sky} = \Omega_{\rm sky} = 13800\times (\pi/180)^2\,\,{\rm rad}^2$, mimicking an LSST-like survey. For the redshift-dependent case (thin-blue curve in Fig. \ref{fig:snr}), we compute the bispectrum using Eq. \ref{eq:bispectrumFeff}, which better represents the effects on non-linearities. For the fixed survey volume, we compute the signal with Eq. \ref{eq:bispectrumFeff} (thick red), and at the tree-level limit (dashed black), i.e. when the functions $\lbrace\tilde{a}, \tilde{b}, \tilde{c}\rbrace \rightarrow 1$.

From the qualitative point of view, there is no known reason to expect an increase in the SNR for realistic situations (presence of foregrounds and instrumental noise for the case of the HI field, or the presence of photo-$z$ errors and shot noise, for the case of galaxies). In fact, the analysis of \citet{cunnington2021} based on simulations shows that foreground removal reduces the bispectrum’s SNR on equilateral configurations by 8\%, while the beam reduces it by 62\%. Since foreground cleaning removes the largest scales, we expect the impact on squeezed configurations to be even larger. Finally, the fingers-of-god effect coming from non-linear redshift-space distortions is another source of damping to the bispectrum for both galaxies and HI, reducing the signal even further. 

\bsp
\label{lastpage}
\end{document}